\documentclass[twocolumn,pra,superscriptaddress,showpacs]{revtex4}
\usepackage{amsmath}
\usepackage{amssymb}
\usepackage{graphicx}
\usepackage{bm}

\begin{document}

\title{The Electromagnetic Green's Function for Layered Topological Insulators}

\author{J. A. Crosse}
\email{alexcrosse@gmail.com}
\affiliation{Department of Electrical and Computer Engineering, National University of Singapore, 4 Engineering Drive 3, Singapore 117583.}
\author{Sebastian Fuchs}
\affiliation{Physikalisches Institut, Albert-Ludwigs-Universit\"at Freiburg, Hermann-Herder-Str. 3, 79104 Freiburg, Germany.}
\author{Stefan Yoshi Buhmann}
\affiliation{Physikalisches Institut, Albert-Ludwigs-Universit\"at Freiburg, Hermann-Herder-Str. 3, 79104 Freiburg, Germany.}
\affiliation{Freiburg Institute for Advanced Studies, Albert-Ludwigs-Universit\"at Freiburg,\\ Albertstra\ss e 19, 79104 Freiburg, Germany.}

\date{\today}

\begin{abstract}
The dyadic Green's function of the inhomogeneous vector Helmholtz equation describes the field pattern of a single frequency point source. It appears in the mathematical description of many areas of electromagnetism and optics including both classical and quantum, linear and nonlinear optics, dispersion forces (such as the Casimir and Casimir-Polder forces) and in the dynamics of trapped atoms and molecules. Here, we compute the Green's function for a layered topological insulator. Via the magnetoelectric effect, topological insulators are able to mix the electric, $\mathbf{E}$, and magnetic induction, $\mathbf{B}$, fields and, hence, one finds that the $TE$ and $TM$ polarizations mix on reflection from/transmission through an interface. This leads to novel field patterns close to the surface of a topological insulator.
\end{abstract}

\pacs{78.20.-e, 78.20.Ek, 78.67.Pt, 42.25.Gy} 

\maketitle

\section{Introduction}

Topological insulators are a class of time-reversal symmetric materials that display non-trivial topological order and are characterized by an insulating bulk with protected conducting edge states \cite{TIrev1,TIrev2}. This type of material was first predicted \cite{2DTI1} and then observed \cite{2DTI2} in $2D$ in HgTe/CdTe quantum wells and subsequently in $3D$ in Group V and Group V/VI alloys that display strong enough spin orbit coupling to induce band inversion - $\mathrm{Bi_{1-x}Sb_{x}}$ in the first instance \cite{3DTI1, 3DTI2} and then in $\mathrm{Bi_{2}Se_{3}}$,  $\mathrm{Bi_{2}Te_{3}}$ and $\mathrm{Sb_{2}Te_{3}}$ \cite{3DTI3,3DTI4} to name but a few examples. Owing to their unusual band structure, these materials display a number of unique electronic properties, the most notable of which is the quantum spin hall effect where quantized surface spin currents are observed even though the usual charge currents are absent \cite{QSHE1,QSHE2}.

In addition to their interesting electronic properties, topological insulators also display a number of unusual electromagnetic properties. Specifically, topological insulators have the ability to mix electric, $\mathbf{E}$, and magnetic induction, $\mathbf{B}$, fields \cite{monopole,TITFT}, a feature which has a pronounced affect on the optical response of the material \cite{chang, grushin1, grushin2}. In particular, this magnetoelectric $\mathbf{E}-\mathbf{B}$ mixing allows one to realise an axionic material \cite{wilczek,obukhov}. Such materials are described by the Lagrangian density $\mathcal{L}_{0} + \mathcal{L}_{axion}$, where $\mathcal{L}_{0}$ is the usual electromagnetic Lagrangian density and $\mathcal{L}_{axion}$ is a term that couples the electric and magnetic induction fields. This additional electromagnetic interaction reads
\begin{equation}
\mathcal{L}_{axion} = \frac{\alpha}{4\pi^{2}}\frac{\Theta(\mathbf{r},\omega)}{\mu_{0}c}\mathbf{E}(\mathbf{r},\omega)\cdot\mathbf{B}(\mathbf{r},\omega),
\label{L}
\end{equation}
where $\alpha$ is the fine structure constant and $\Theta(\mathbf{r},\omega)$ is termed the axion field in particle physics (although, as far as electromagnetism is concerned, it merely acts as a space and frequency dependent coupling parameter). In order to realise such a material in a topological insulator, a time symmetry breaking perturbation of sufficient size must be introduced to the surface to induce a gap, thereby converting the material into a full insulator. Such a gap can be opened by introducing ferromagnetic dopants to the surface (12\% Fe doping in $\mathrm{Bi_{2}Se_{3}}$ leads to a mid-infrared gap of 50meV/25$\mu$m \cite{chen}) or by the application of an external static magnetic field \cite{qi}. In such a time-reversal-symmetry-broken topological insulator (TSB-TI) the constitutive relations are altered and, hence, the optical properties of the material change dramatically.

Here, we derive the electromagnetic Green's function for a layered TSB-TI. The electromagnetic Green's function is the solution to the vector Helmholtz equation for a single frequency point source and can be used to generate general field solutions for an arbitrary distribution of sources. This function has a wide range of applications in both classical \cite{chew, jackson} and quantum optics \cite{dung, acta, buhmann,Buhmann12c} and is an important component in studies of linear \cite{IO} and non-linear \cite{nonlinearH,photon} optics, Casimir \cite{casimir} and Casmir-Polder \cite{cp1, cp2} forces, decoherence \cite{decoherence} and the dynamics of trapped atoms \cite{atom} and molecules \cite{molecule1, molecule2}. Thus, knowledge of the Green's function is of value to a great many fields.

\section{Maxwell Equations}

As with all electromagnetic studies, we begin with the Maxwell equations and constitutive relations for the material in question. For a TSB-TI these are \cite{TITFT, chang, obukhov}
\begin{gather}
\bm{\nabla}\cdot\mathbf{B}(\mathbf{r},\omega) = 0,\label{max1}\\
\bm{\nabla}\times\mathbf{E}(\mathbf{r},\omega) - i\omega\mathbf{B}(\mathbf{r},\omega) = \mathbf{0}, \label{max2}\\
\bm{\nabla}\cdot\mathbf{D}(\mathbf{r},\omega) = \rho(\mathbf{r},\omega),\label{max3}\\
\bm{\nabla}\times\mathbf{H}(\mathbf{r},\omega) + i\omega\mathbf{D}(\mathbf{r},\omega) = \mathbf{J}(\mathbf{r},\omega),\label{max4} 
\end{gather}
and
\begin{align}
\mathbf{D}(\mathbf{r},\omega) &= \varepsilon_0\varepsilon(\mathbf{r},\omega)\mathbf{E}(\mathbf{r},\omega)\nonumber\\
&\qquad + \frac{\alpha}{\pi}\frac{\Theta(\mathbf{r},\omega)}{\mu_{0}c}\mathbf{B}(\mathbf{r},\omega)+\mathbf{P}_{N}(\mathbf{r},\omega),\label{con1} \\
\mathbf{H}(\mathbf{r},\omega) &= \frac{1}{\mu_0\mu(\mathbf{r},\omega)}\mathbf{B}(\mathbf{r},\omega)\nonumber\\
&\qquad - \frac{\alpha}{\pi}\frac{\Theta(\mathbf{r},\omega)}{\mu_{0}c}\mathbf{E}(\mathbf{r},\omega)-\mathbf{M}_{N}(\mathbf{r},\omega),\label{con2} 
\end{align}
where $\alpha$ is the fine structure constant and $\varepsilon(\mathbf{r},\omega)$, $\mu(\mathbf{r},\omega)$ and $\Theta(\mathbf{r},\omega)$ are the dielectric permittivity, magnetic permeability and axion coupling respectively, the latter of which takes even multiples of $\pi$ in a conventional magneto-dielectric and odd multiples of $\pi$ in TSB-TI, with the magnitude and sign of the multiple given by the strength and direction of the time symmetry breaking perturbation. The $\mathbf{P}_{N}(\mathbf{r},\omega)$ and $\mathbf{M}_{N}(\mathbf{r},\omega)$ terms are the noise polarization and magnetization, respectively. These terms are Langevin noise terms that model absorption within the material \cite{acta}. These relations can be derived from the Lagrangian density in Eq. \eqref{L} \cite{wilczek}. Using the above constitutive relations, one can show that the frequency components of the electric field obey the inhomogeneous Helmholtz equation
\begin{multline}
\bm{\nabla}\times\frac{1}{\mu(\mathbf{r},\omega)}\bm{\nabla}\times\mathbf{E}(\mathbf{r},\omega)-\frac{\omega^2}{c^2}\varepsilon(\mathbf{r},\omega)\mathbf{E}(\mathbf{r},\omega)\\ 
-i\frac{\omega}{c}\frac{\alpha}{\pi}\left[\bm{\nabla}\Theta(\mathbf{r},\omega)\times\mathbf{E}(\mathbf{r},\omega)\right]\\
= i\omega\mu_{0}\left[\mathbf{J}_{E}(\mathbf{r},\omega) + \mathbf{J}_{N}(\mathbf{r},\omega)\right],
\label{EHelmholtz}
\end{multline}
where $\mathbf{J}_{E}(\mathbf{r},\omega)$ is the source term for electromagnetic waves generated by external currents and $\mathbf{J}_{N}(\mathbf{r},\omega) = -i\omega\mathbf{P}_{N}(\mathbf{r},\omega) + \bm{\nabla}\times\mathbf{M}_{N}(\mathbf{r},\omega)$ is the source term for electromagnetic waves generated by noise fluctuations within the material. If the axion coupling is homogeneous, $\Theta(\mathbf{r},\omega) = \Theta(\omega)$, then the last term on the left-hand side vanishes and one finds that the propagation of the electric field is the same as in a conventional magneto-dielectric. As a result, electromagnetic waves propagating within a homogeneous TSB-TI retain there usual properties - dispersion is linear, the phase and group velocities are proportional to the usual refractive index, the fields are transverse and orthogonal polarizations do not mix. Thus, the effects of the axion coupling are only felt when the axion coupling varies in space. For layered, homogeneous media this will occur only at the interfaces where the properties of the medium change.

\section{Fresnel Coefficients}

An important set of functions for any layered media are the Fresnel coefficients for reflection and transmission at each interface. These functions are required to construct the Green's function. In fact, the Fresnel coefficients for TSB-TIs have been studied before \cite{chang,obukhov}, however, the standard expression for the Green's function requires a slightly different form for the coefficients compared to those in previous work \cite{chew,buhmann}. Furthermore, certain aspects of TSB-TIs mean that the usual method of computing this form of the coefficients leads to incorrect results. For these reasons, it is worth revisiting the derivation in some detail.

\begin{figure}[b]
\centering
\includegraphics[width=0.8\linewidth]{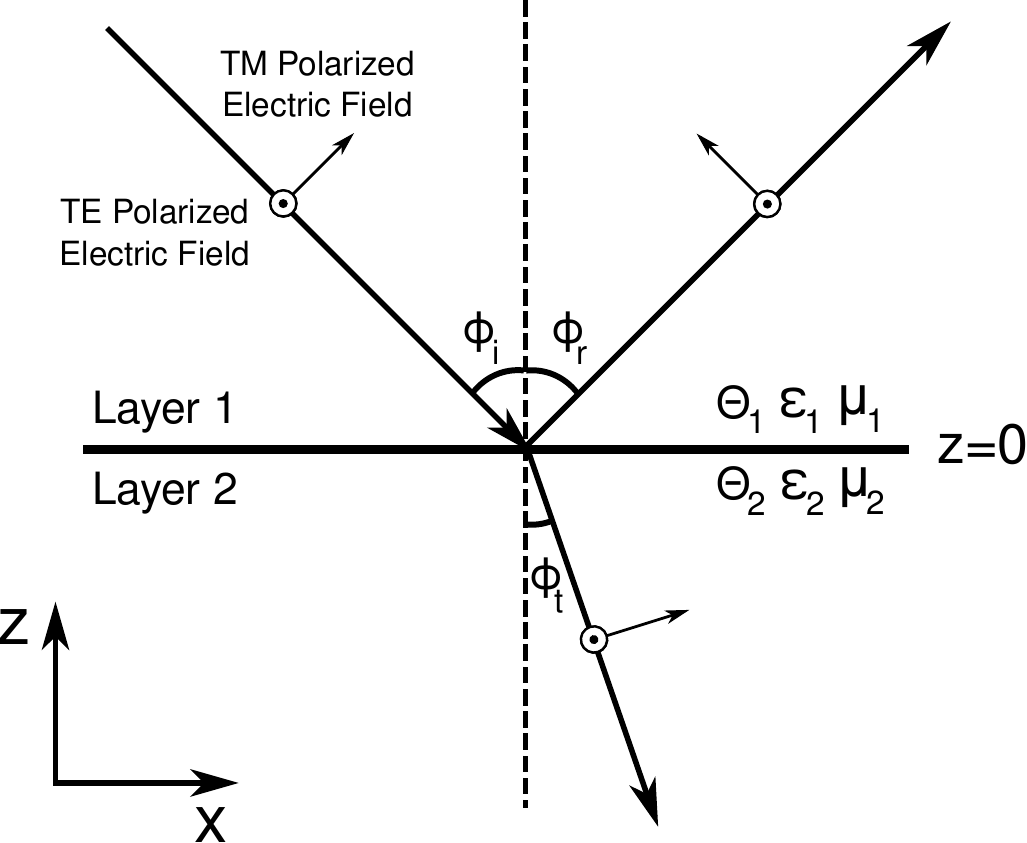}
\caption{The interface between two topological insulators.}
\label{interface}
\end{figure}

In the derivation of the Fresnel coefficients for a conventional magneto-dielectric one usually defines two polarizations; the $TE$ polarization, where the electric field, $\mathbf{E}$, is parallel to the interface, and the $TM$ polarization, where the magnetic field, $\mathbf{H}$, is parallel to the interface. Since, from Eq. \eqref{EHelmholtz}, the electric field propagation is unaffected by a homogeneous axion coupling one can see that the $TE$ polarization is unchanged. However, from Eq. \eqref{con2}, one can see that the magnetic field, $\mathbf{H}$, is no longer perpendicular to the electric field, $\mathbf{E}$. Thus defining the $TM$ polarization in terms of the magnetic field, $\mathbf{H}$, leads to two polarizations that are not orthogonal and hence incorrect expressions for the Fresnel coefficients. Furthermore, we would expect the two polarizations to mix at the interface via the magnetoelectric coupling, hence defining the two polarizations in terms of different fields leads to awkward expressions. The simplest approach is to work solely with the electric field, $\mathbf{E}$. Thus, for media layered in the $\hat{z}$ direction and light incident in the $x-z$ plane, we define the $TE$ polarization as the polarization with $E_{y} \neq 0$ and $E_{x} = 0$, $E_{z} = 0$ and the $TM$ polarization as the polarization with $E_{y} = 0$ and $E_{x} \neq  0$, $E_{z} \neq 0$, [See Fig. \ref{interface}].

We proceed by considering waves of a specific ($TE$, $TM$) polarization incident on an interface between two homogeneous isotropic TSB-TIs. By matching the waves in each half-space using the electromagnetic jump conditions
\begin{gather}
\hat{z}\times\mathbf{E}_{1} = \hat{z}\times\mathbf{E}_{2},\label{jump1}\\
\hat{z}\times\mathbf{H}_{1} = \hat{z}\times\mathbf{H}_{2},\label{jump2}
\end{gather}
which relate the transverse components of the electric and magnetic fields on either side of the interface, the Fresnel coefficients can be found.

First we consider a $TE$ polarized plane wave incident on the interface from layer $1$ [See Fig. \ref{interface}]. The electric field ansatz for each region is 
\begin{align}
E_{x,1} &= -E_{0}\frac{k_{z,1}}{k_{1}}e^{ik_{z,1}z+ik_{p}x}R_{TM,TE},\\
E_{y,1} &= -E_{0}\left[e^{-ik_{z,1}z+ik_{p}x}+e^{ik_{z,1}z+ik_{p}x}R_{TE,TE}\right],\\
E_{z,1} &= E_{0}\frac{k_{p}}{k_{1}}e^{ik_{z,1}z+ik_{p}x}R_{TM,TE},\\
E_{x,2} &= E_{0}\frac{k_{z,2}}{k_{2}}e^{-ik_{z,2}z+ik_{p}x}T_{TM,TE},\\
E_{y,2} &= -E_{0}e^{-ik_{z,2}z+ik_{p}x}T_{TE,TE},\\
E_{z,2} &= E_{0}\frac{k_{p}}{k_{2}}e^{-ik_{z,2}z+ik_{p}x}T_{TM,TE},
\end{align}
where the $k_{z}/k_{1}=\cos\phi_{r}$, $k_{p}/k_{1}=\sin\phi_{r}$, $k_{z}/k_{2}=\cos\phi_{t}$, $k_{p}/k_{2}=\sin\phi_{t}$. It has been previously shown that Snell's law holds for TSB-TI's so $\phi_{i}=\phi_{r}$ \cite{chang,obukhov}. From Eqs. \eqref{max2} and \eqref{con2} we obtain
\begin{align}
H_{x,1} &= -E_{0}\frac{k_{z,1}}{\mu_{0}\mu_{1}\omega}\left[e^{-ik_{z,1}z+ik_{p}x}-e^{ik_{z,1}z+ik_{p}x}R_{TE,TE}\right]\nonumber\\
& +E_{0}\frac{\alpha}{\pi}\frac{\Theta_{1}}{\mu_{0}c}\frac{k_{z,1}}{k_{1}}e^{ik_{z,1}z+ik_{p}x}R_{TM,TE},\\
H_{y,1} &= -E_{0}\frac{k_{1}}{\mu_{0}\mu_{1}\omega}e^{ik_{z,1}z+ik_{p}x}R_{TM,TE}\nonumber\\
& +E_{0}\frac{\alpha}{\pi}\frac{\Theta_{1}}{\mu_{0}c}\left[e^{-ik_{z,1}z+ik_{p}x}+e^{ik_{z,1}z+ik_{p}x}R_{TE,TE}\right],\\
H_{x,2} &= -E_{0}\frac{k_{z,2}}{\mu_{0}\mu_{2}\omega}e^{-ik_{z,1}z+ik_{p}x}T_{TE,TE}\nonumber\\
& -E_{0}\frac{\alpha}{\pi}\frac{\Theta_{2}}{\mu_{0}c}\frac{k_{z,2}}{k_{2}}e^{-ik_{z,1}z+ik_{p}x}T_{TM,TE},\\
H_{y,2} &= -E_{0}\frac{k_{2}}{\mu_{0}\mu_{2}\omega}e^{-ik_{z,1}z+ik_{p}x}T_{TM,TE}\nonumber\\
& +E_{0}\frac{\alpha}{\pi}\frac{\Theta_{2}}{\mu_{0}c}e^{-ik_{z,1}z+ik_{p}x}T_{TE,TE}.
\end{align}
From Eqs. \eqref{jump1} and \eqref{jump2} we can find the boundary conditions for the fields at the interface at $z=0$
\begin{gather}
1+R_{TE,TE} = T_{TE,TE},\\
-\frac{k_{z,1}}{n_{1}}R_{TM,TE} = \frac{k_{z,2}}{n_{2}}T_{TM,TE},\\
-\frac{k_{z,1}}{\mu_{1}}\left[1-R_{TE,TE}\right] + \frac{\alpha}{\pi}\Theta_{1}\frac{k_{z,1}}{n_{1}}R_{TM,TE}\qquad\qquad\nonumber\\
\qquad\qquad =- \frac{k_{z,2}}{\mu_{2}}T_{TE,TE} - \frac{\alpha}{\pi}\Theta_{2}\frac{k_{z,2}}{n_{2}}T_{TM,TE},\\
\frac{n_{1}}{\mu_{1}}R_{TM,TE} - \frac{\alpha}{\pi}\Theta_{1}\left[1+R_{TE,TE}\right]\qquad\qquad\nonumber\\
\qquad\qquad = \frac{n_{2}}{\mu_{2}}T_{TM,TE} - \frac{\alpha}{\pi}\Theta_{2}T_{TE,TE},
\end{gather}
where we have used the dispersion relation $k=n\omega/c$. Solving the above system of equations gives
\begin{align}
R_{TE,TE} &= \frac{(\mu_{2}k_{z,1}-\mu_{1}k_{z,2})\Omega_{\varepsilon}-k_{z,1}k_{z,2}\Delta^{2}}{(\mu_{2}k_{z,1}+\mu_{1}k_{z,2})\Omega_{\varepsilon}+k_{z,1}k_{z,2}\Delta^{2}},\\ 
R_{TM,TE} &= \frac{-2\mu_{2}n_{1}k_{z,1}k_{z,2}\Delta}{(\mu_{2}k_{z,1}+\mu_{1}k_{z,2})\Omega_{\varepsilon}+k_{z,1}k_{z,2}\Delta^{2}},\\ 
T_{TE,TE} &= \frac{2\mu_{2}k_{z,1}\Omega_{\varepsilon}}{(\mu_{2}k_{z,1}+\mu_{1}k_{z,2})\Omega_{\varepsilon}+k_{z,1}k_{z,2}\Delta^{2}},\\
T_{TM,TE} &= \frac{2\mu_{2}n_{2}k_{z,1}^{2}\Delta}{(\mu_{2}k_{z,1}+\mu_{1}k_{z,2})\Omega_{\varepsilon}+k_{z,1}k_{z,2}\Delta^{2}},
\end{align}
where $\Delta = \alpha\mu_{1}\mu_{2}(\Theta_{2}-\Theta_{1})/\pi$ and $\Omega_{\varepsilon} = \mu_{1}\mu_{2}(k_{z,1}\varepsilon_{2}+k_{z,2}\varepsilon_{1})$, with the factors of $\varepsilon$ appearing via the definition of the refractive index, $n^{2} = \mu\varepsilon$. Note that when $\Theta_{2} - \Theta_{1}\rightarrow 0$ (i.e when the axion couplings vanish or when they are the same across the interface), $R_{TM,TE}, T_{TM,TE} \rightarrow 0$ and $R_{TE,TE}$ and $T_{TE,TE}$ reduce to the usual reflection coefficients for normal magneto-electric materials \cite{chew, buhmann}.

Next, we consider a $TM$ polarized plane wave incident on the interface from layer $1$ [See Fig. \ref{interface}]. The electric field ansatz for each region is now
\begin{align}
E_{x,1} &= E_{0}\frac{k_{z,1}}{k_{1}}\left[e^{-ik_{z,1}z+ik_{p}x}-e^{ik_{z,1}z+ik_{p}x}R_{TM,TM}\right],\\
E_{y,1} &= -E_{0}e^{ik_{z,1}z+ik_{p}x}R_{TE,TM},\\
E_{z,1} &= E_{0}\frac{k_{p}}{k_{1}}\left[e^{-ik_{z,1}z+ik_{p}x}+e^{ik_{z,1}z+ik_{p}x}R_{TM,TM}\right],\\
E_{x,2} &= E_{0}\frac{k_{z,2}}{k_{2}}e^{-ik_{z,2}z+ik_{p}x}T_{TM,TM},\\
E_{y,2} &= -E_{0}e^{-ik_{z,2}z+ik_{p}x}T_{TE,TM},\\
E_{z,2} &= E_{0}\frac{k_{p}}{k_{2}}e^{-ik_{z,2}z+ik_{p}x}T_{TM,TM},
\end{align}
where, again, components of the wavenumber are related to the angles of incidence, reflection and transmission and Snells law holds. From Eqs. \eqref{max2} and Eq. \eqref{con2} we obtain
\begin{align}
H_{x,1} &= E_{0}\frac{k_{z,1}}{\mu_0\mu_{1}\omega}e^{ik_{z,1}z+ik_{p}x}R_{TE,TM}\nonumber\\
& - E_{0}\frac{\alpha}{\pi}\frac{\Theta_{1}}{\mu_{0}c}\frac{k_{z,1}}{k_{1}}\left[e^{-ik_{z,1}z+ik_{p}x}-e^{ik_{z,1}z+ik_{p}x}R_{TM,TM}\right],\\
H_{y,1} &= -E_{0}\frac{k_{1}}{\mu_0\mu_{1}\omega}\left[e^{-ik_{z,1}z+ik_{p}x}+e^{ik_{z,1}z+ik_{p}x}R_{TM,TM}\right]\nonumber\\
& + E_{0}\frac{\alpha}{\pi}\frac{\Theta_{1}}{\mu_{0}c}e^{ik_{z,1}z+ik_{p}x}R_{TE,TM},\\
H_{x,2} &= -E_{0}\frac{k_{z,2}}{\mu_0\mu_{2}\omega}e^{-ik_{z,2}z+ik_{p}x}T_{TE,TM}\nonumber\\
& - E_{0}\frac{\alpha}{\pi}\frac{\Theta_{2}}{\mu_{0}c}\frac{k_{z,2}}{k_{2}}e^{-ik_{z,2}z+ik_{p}x}T_{TM,TM},\\
H_{y,2} &= -E_{0}\frac{k_{2}}{\mu_0\mu_{2}\omega}e^{-ik_{z,2}z+ik_{p}x}T_{TM,TM}\nonumber\\
& +E_{0}\frac{\alpha}{\pi}\frac{\Theta_{2}}{\mu_{0}c}e^{-ik_{z,2}z+ik_{p}x}T_{TE,TM}.
\end{align}
From Eqs. \eqref{jump1} and \eqref{jump2} we can, again, find the boundary conditions for the fields at the interface at $z=0$
\begin{gather}
\frac{k_{z,1}}{n_{1}}\left[1-R_{TM,TM}\right] = \frac{k_{z,2}}{n_{2}}T_{TM,TM},\\
R_{TE,TM} = T_{TE,TM},\\
\frac{k_{z,1}}{\mu_{1}}R_{TE,TM} - \frac{\alpha}{\pi}\Theta_{1}\frac{k_{z,1}}{n_{1}}\left[1-R_{TM,TM}\right] \qquad\qquad\nonumber\\
\qquad\qquad = - \frac{k_{z,2}}{\mu_{2}}T_{TE,TM} - \frac{\alpha}{\pi}\Theta_{2}\frac{k_{z,2}}{n_{2}}T_{TM,TM},\\
\frac{n_{1}}{\mu_{1}}\left[1+R_{TM,TM}\right] - \frac{\alpha}{\pi}\Theta_{1}R_{TE,TM} \qquad\qquad\nonumber\\
\qquad\qquad = \frac{n_{2}}{\mu_{2}}T_{TM,TM} - \frac{\alpha}{\pi}\Theta_{2}T_{TE,TM},
\end{gather}
where, once more, $k=n\omega/c$ has been used. Solving the above system of equations gives
\begin{align}
R_{TM,TM} &= \frac{(\varepsilon_{2}k_{z,1}-\varepsilon_{1}k_{z,2})\Omega_{\mu}+ k_{z,1}k_{z,2}\Delta^{2}}{(\varepsilon_{2}k_{z,1}+\varepsilon_{1}k_{z,2})\Omega_{\mu} +k_{z,1}k_{z,2}\Delta^{2}},\\ 
R_{TE,TM} &= \frac{-2\mu_{2}n_{1}k_{z,1}k_{z,2}\Delta}{(\varepsilon_{2}k_{z,1}+\varepsilon_{1}k_{z,2})\Omega_{\mu} +k_{z,1}k_{z,2}\Delta^{2}},\\ 
T_{TM,TM} &= \frac{n_{2}}{n_{1}}\frac{2\varepsilon_{1}k_{z,1}\Omega_{\mu}}{(\varepsilon_{2}k_{z,1}+\varepsilon_{1}k_{z,2})\Omega_{\mu} +k_{z,1}k_{z,2}\Delta^{2}},\\
T_{TE,TM}  &=  \frac{-2\mu_{2}n_{1}k_{z,1}k_{z,2}\Delta}{(\varepsilon_{2}k_{z,1}+\varepsilon_{1}k_{z,2})\Omega_{\mu} +k_{z,1}k_{z,2}\Delta^{2}},
\end{align}
where $\Delta = \alpha\mu_{1}\mu_{2}(\Theta_{2}-\Theta_{1})/\pi$ and $\Omega_{\mu} = \mu_{1}\mu_{2}(k_{z,1}\mu_{2}+k_{z,2}\mu_{1})$, with the factors of $\varepsilon$, again, appearing via the definition of the refractive index. Once again, when $\Theta_{2} - \Theta_{1}\rightarrow 0$ (i.e when the axion couplings vanish or when they are the same across the interface), $R_{TE,TM}, T_{TE,TM} \rightarrow 0$ and $R_{TM,TM}$ and $T_{TM,TM}$ reduce to the usual reflection coefficients for normal magneto-electric materials \cite{chew, buhmann}. (Note that in \cite{chew}, unlike \cite{buhmann}, the transmission coefficient differs from the above result by the ratio of the impedances of the two layers. This is because the $TM$ coefficients are derived using the $\mathbf{H}$-field instead of the $\mathbf{E}$-field.) 

\begin{figure}[h]
\centering
\includegraphics[width=0.73\linewidth]{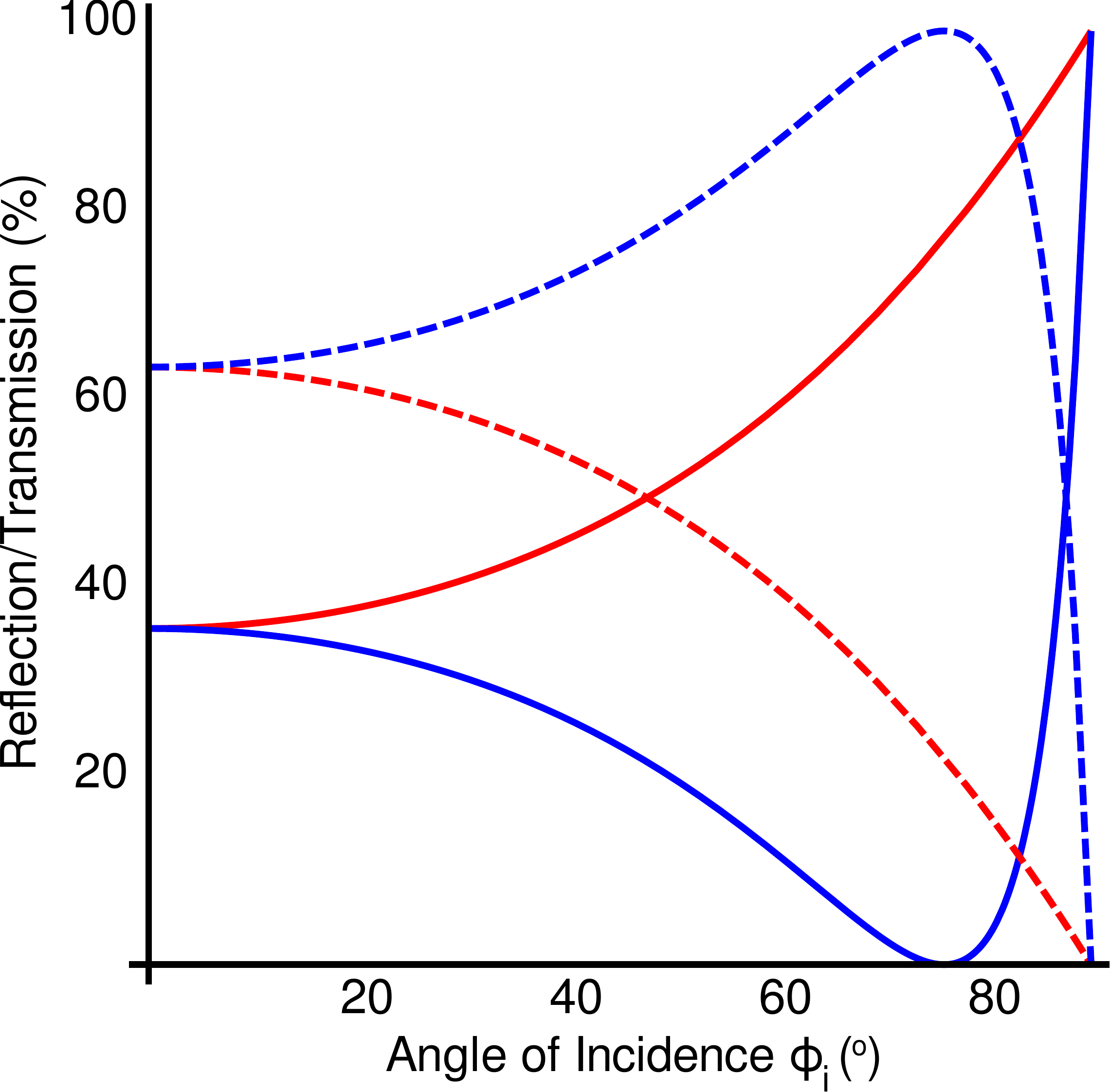}
\caption{(Color online) The \% reflection (solid) and transmission (dashed) for $TE$ (red) and $TM$ (blue) polarized waves at a vacuum- (layer $1$) magneto-dielectic- (layer $2$) interface as a function of the incident angle, $\phi_{i}$. Here, $\mu_{1} = \mu_{2} = 1$, $\varepsilon_{1} = 1$ and $\varepsilon_{2} = 16$.}
\label{FMEM}
\end{figure}

\begin{figure}[h]
\centering
\includegraphics[width=0.8\linewidth]{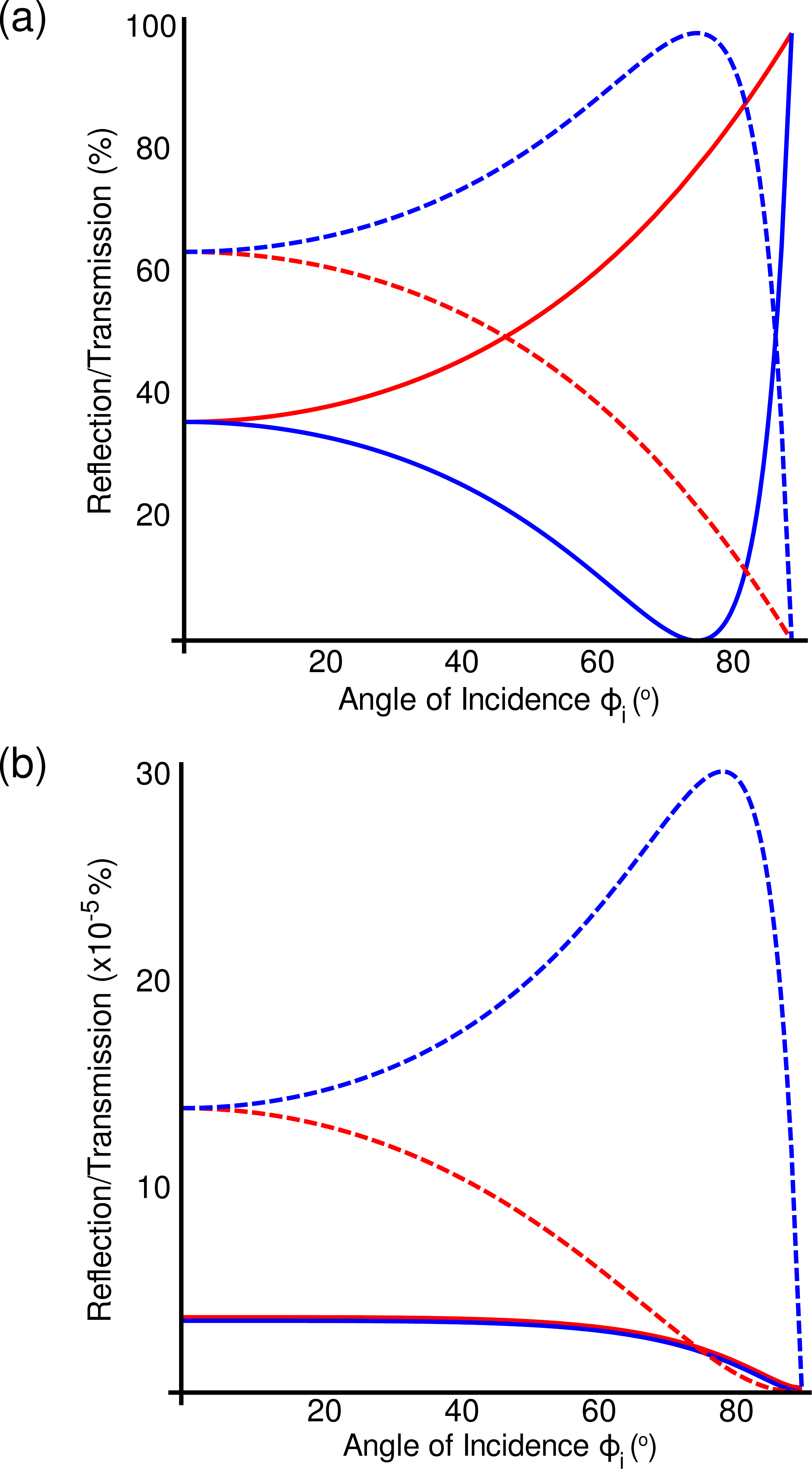}
\caption{(Color online) (a) The \% reflection (solid) and transmission (dashed) for $TE$ (red) and $TM$ (blue) polarized waves at a vacuum- (layer $1$) TRSB-TI- (layer $2$) interface as a function of the incident angle, $\phi_{i}$. (b) The \% reflection (solid) and transmission (dashed) for $TE \rightarrow TM$ mixing (red) and $TM \rightarrow TE$ mixing (blue) at a vacuum- (layer $1$) TRSB-TI- (layer $2$) interface as a function of the incident angle, $\phi_{i}$. In both cases, $\mu_{1} = \mu_{2} = 1$, $\varepsilon_{1} = 1$, $\varepsilon_{2} = 16$, $\Theta_{1} = 0$ and $\Theta_{2} = \pi$.}
\label{FTI}
\end{figure}

The Fresnel coefficients for the energy flux can be found by comparing the $\hat{z}$-component of the Poynting vector, $\mathbf{S} = \mathbf{E}\times\mathbf{H}$, on each side of the interface. One finds that they are related to the above field coefficients via
\begin{gather}
r_{i,j} = |R_{i,j}|^{2},\\
t_{i,j} = \frac{k_{z,2}}{k_{z,1}}\frac{\mu_{1}}{\mu_{2}}|T_{i,j}|^{2},
\end{gather}
where $i, j \in TE, TM$ and the prefactor in the transmission coefficients accounting for the change in flux area as the field passes through the interface. Figure \ref{FMEM} shows the reflection, $r_{i,i}$, and transmission, $t_{i,i}$, as a function of incident angle for a $600$nm optical plane wave, incident from the vacuum, encountering a conventional magneto-dielectric with $\varepsilon = 16$ and $\mu = 1$ (such values are similar to those for $\mathrm{Bi_{2}Se_{3}}$ at high frequencies \cite{bi2se3}). 
In this case the mixing coefficients vanish and the polarization state of the incident light is preserved by the interface. It is easy to see that for incident $TE$ polarized light $r_{TE,TE}+t_{TE,TE}=1$ hence the $TE$ energy flux is preserved at the interface (a similar expression holds for $TM$ polarized light). In comparison, Fig.~\ref{FTI} shows the reflection, $r_{i,i}$, transmission, $t_{i,i}$, and mixing, $r_{i,j}$/$t_{i,j}$ ($i\neq j$), as a function of incident angle for a plane wave of similar wavelength, incident from the vacuum, encountering a TSB-TI with $\varepsilon = 16$, $\mu = 1$ and $\Theta_{2} = \pi$.   
In this case the mixing coefficients are non-zero and the polarization states of the incident light mix at the interface. Finally, one can show that for incident $TE$ polarized light, $r_{TE,TE}+r_{TM,TE}+t_{TE,TE}+t_{TM,TE}=1$ hence $TE$ energy flux is still preserved at the interface (again a similar expression holds for $TM$ polarized light).

In order to better understand the mixing coefficients, it is informative to look at the case of a pure TSB-TI where the permittivity and permeability are that of the vacuum and only the axion coupling changes on the interface. This allows one to remove the magneto-dielectric effects from the system and isolate the effect of the axion coupling. In this case $k_{z,1} = k_{z,2}$ and the reflection and transmission coefficients reduce to
{\allowdisplaybreaks
\begin{gather}
R_{TE,TE} = -R_{TM,TM} = \frac{-\Delta^{2}}{4+\Delta^{2}},\label{pureti1}\\ 
T_{TE,TE} = T_{TM,TM} = \frac{4}{4+\Delta^{2}},\\
R_{TM,TE} = R_{TE,TM} = -T_{TM,TE} = T_{TE,TM} = \frac{-2\Delta}{4+\Delta^{2}}\label{pureti3}.
\end{gather}
}
(Similar expressions were found in \cite{obukhov}.) One can see that in the pure TSB-TI limit the reflection and transmission coefficients are no longer a function of incident angle and, hence, the angular dependence of the coefficients is a result of the magneto-dielectric properties of the material rather than the axionic properties. As $\Delta \approx \alpha$, the energy flux reflection and transmission coefficients are $r_{i,i} \approx \alpha^{4}/16 \approx 10^{-10}$ and $t_{i,i} \approx 1$ respectively. Thus one sees near perfect transmission. However, since the axion coupling changes on the interface one still has mixing, the magnitude of which is equal in transmission and reflection $r_{i,j} = t_{i,j} \approx \alpha^{2}/4 \approx 10^{-5}$. 

As a slight diversion we briefly consider reflection, transmission and mixing for large values of $\Delta$. Although such large values are probably not realizable with a topological insulator, this limit is useful in understanding the affect of the axion coupling and may bear some relation to treatments of the axion coupling as a fundamental field. 
\begin{figure}[b]
\centering
\includegraphics[width=0.73\linewidth]{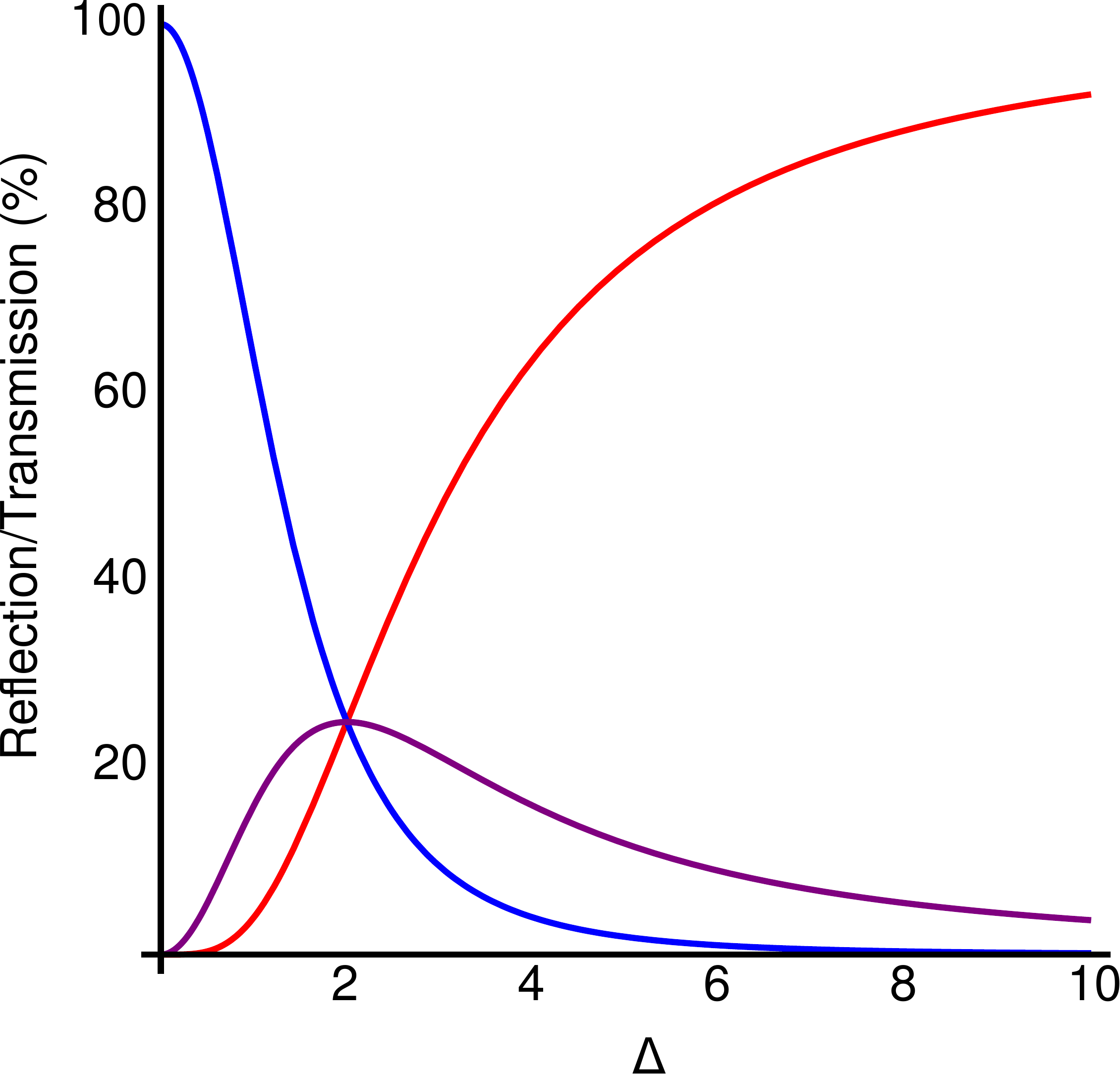}
\caption{(Color online) The \% reflection (red) and transmission (blue) and mixing (purple) as a function of $\Delta$ for a pure TI, with $\mu_{1} = \mu_{2} = \varepsilon_{1} = \varepsilon_{2} = 1$.}
\label{PFTI}
\end{figure}
Figure \ref{PFTI} shows the reflective, $r_{i,i}$, transmissive, $t_{i,i}$, and mixing, $r_{i,j}$, /$t_{i,j}$,  coefficients for the expressions in Eqs. \eqref{pureti1} - \eqref{pureti3} as a function of $\Delta$. For vanishing $\Delta$ one sees that the reflection and mixing coefficients vanish and one has perfect transmission. For $\Delta \rightarrow \infty$, the transmission and mixing coefficients vanish and one approaches a perfect mirror. Thus, for large changes in the axion coupling, the interface becomes purely reflective and no mixing occurs. Further, one sees that the maximum mixing occurs when the reflection and transmission coefficients are equal. 
\begin{figure}[t]
\centering
\includegraphics[width=0.83\linewidth]{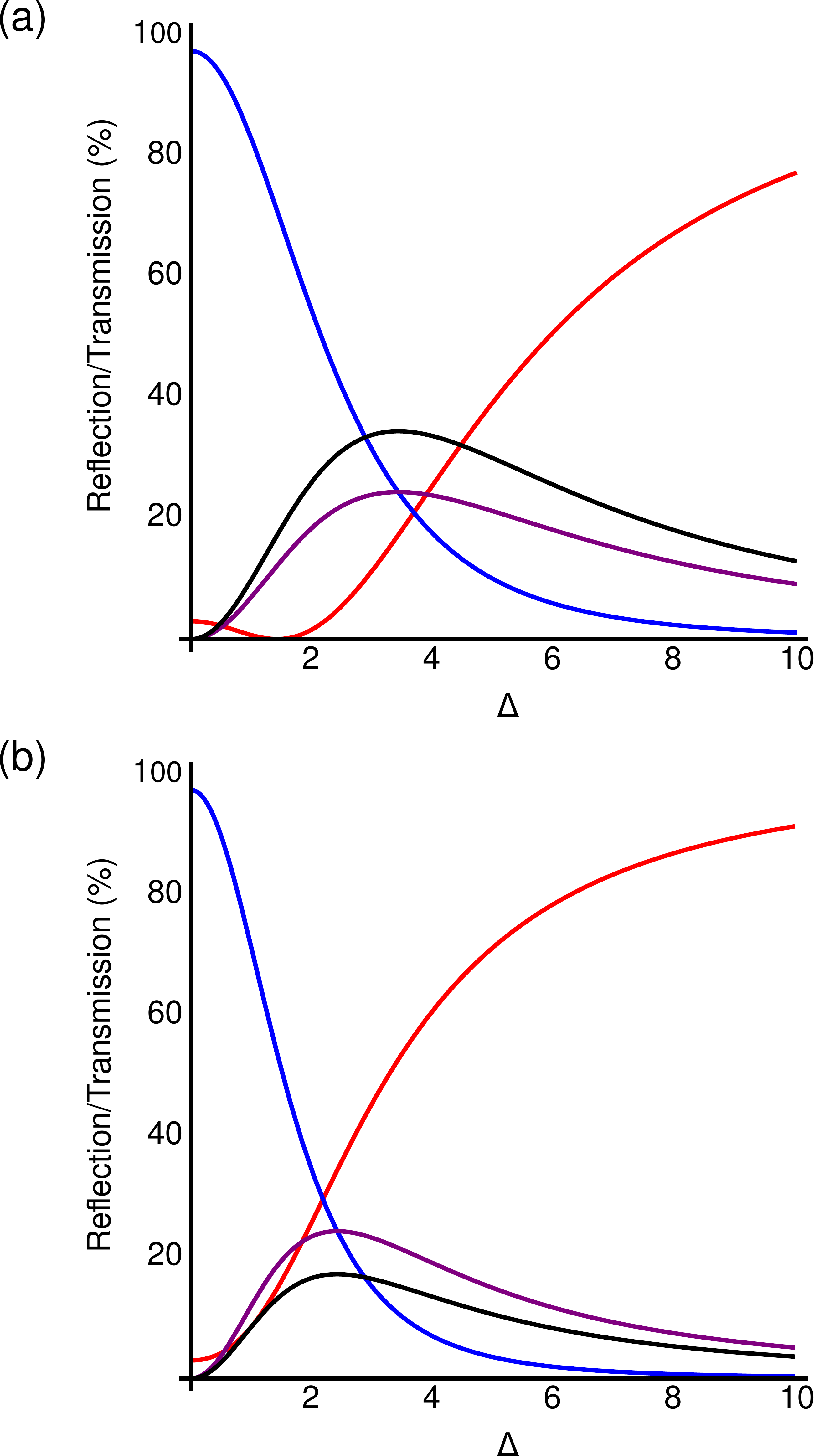}
\caption{(Color online) The \% reflection (red), transmission (blue), reflective mixing (black) and transmissive mixing (purple) as a function of $\Delta$ for a TSB-TI, with (a) $\mu_{1} = \varepsilon_{1} = \varepsilon_{2} = 1$ and $\mu_{2} = 2$ and (b) $\mu_{1} = \mu_{2} = \varepsilon_{2} = 1$ and $\varepsilon_{2} = 2$ for normal incidence angle (hence $TE$ and $TM$ polarizations are indistinguishable).}
\label{PFTI2}
\end{figure}
However, if one changes the relative impedances of the layers, this maximum is shifted. Increasing the impedance (increasing $\mu$ relative to $\varepsilon$) increases the mixing and shifts the peak to $\Delta$ values lower than the reflection-transmission crossing point [See Fig. \ref{PFTI2} (a)], while lowering the impedance (increasing $\varepsilon$ relative to $\mu$) leads to a decrease in the mixing and shifts the peak to $\Delta$ values larger than the reflection-transmission crossing point [See Fig. \ref{PFTI2} (b)]. Thus, we see the mixing is enhanced by the magnetic response of the material and suppressed by the electric response.

Finally, it will be convenient for the rest of this study to write the Fresnel coefficients in matrix form
\begin{gather}
\bm{\underline{R}} = \left(\begin{array}{cc}
R_{TE,TE} & R_{TE,TM} \\
R_{TM,TE} & R_{TM,TM}
\end{array}\right),\\
\bm{\underline{T}} = \left(\begin{array}{cc}
T_{TE,TE} & T_{TE,TM} \\
T_{TM,TE} & T_{TM,TM}
\end{array}\right).
\end{gather}
These matrix transformations act on the field vector $\mathbf{\underline{E}}$ whose components refer to the $TE$ and $TM$ polarizations respectively.

\section{The Generalized Fresnel Coefficients for Multilayered Media}

\begin{figure}[b]
\centering
\includegraphics[width=0.8\linewidth]{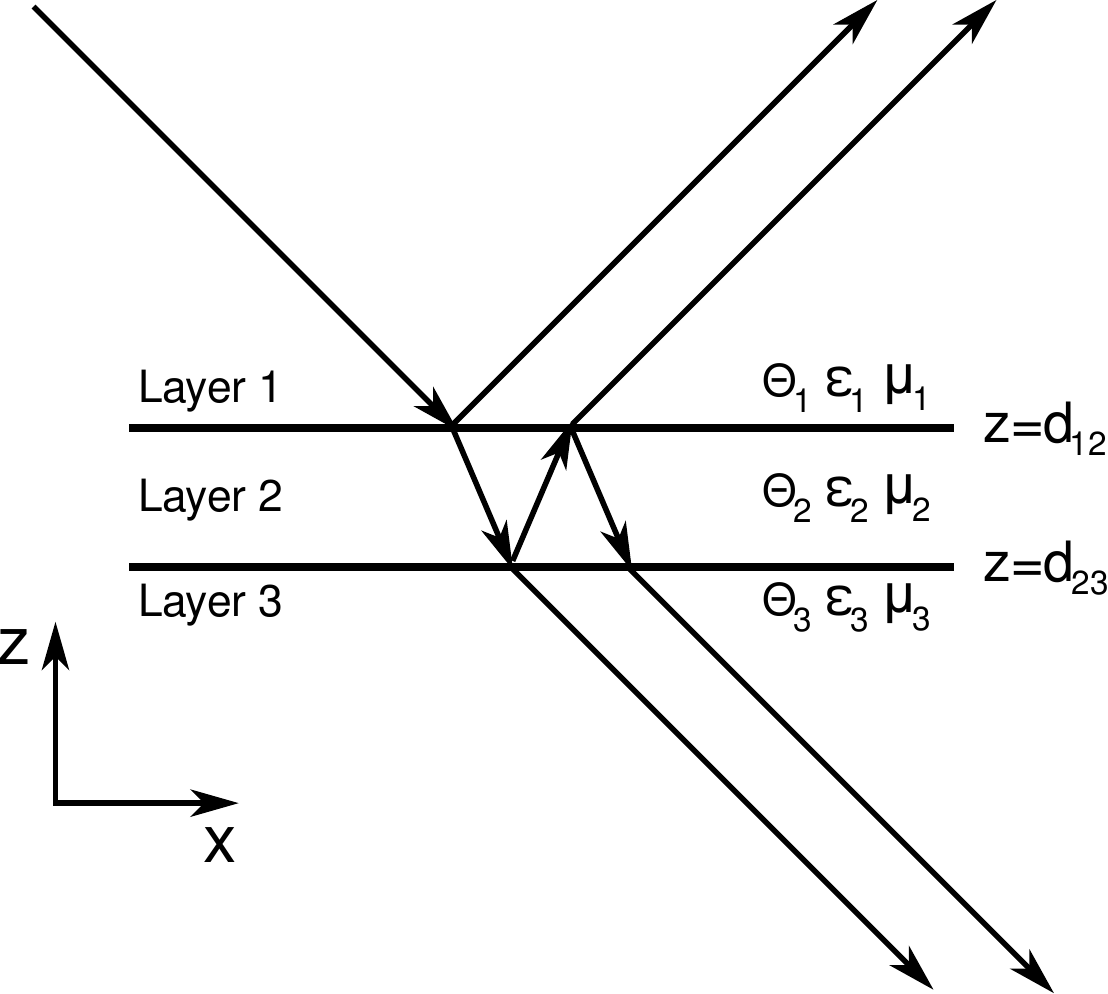}
\caption{A three layered medium.}
\end{figure}
To find the reflection and transmission coefficients for multi-layered media we follow the method of Ref. \cite{chew} and first consider a three layered medium. The wave in layer 1 can be written, in the $TE/TM$ basis as
\begin{equation}
\mathbf{E}_{1} = \left[e^{-ik_{z,1}z}\bm{\underline{I}} + e^{ik_{z,1}(z-2d_{12})}\tilde{\bm{\underline{R}}}_{12}\right]\cdot\mathbf{E}_{0,1},
\end{equation}
where $\bm{\underline{I}}$ is the unit matrix and $\tilde{\bm{\underline{R}}}_{12}$ is the generalized reflection matrix at the layer $1$ - layer $2$ interface, whose form is unknown. Similarly in layer 2 we have
\begin{equation}
\mathbf{E}_{2} = \left[e^{-ik_{z,2}z}\bm{\underline{I}} + e^{ik_{z,1}(z-2d_{23})}\bm{\underline{R}}_{23}\right]\cdot\mathbf{E}_{0,2},
\end{equation}
and in layer 3 we have
\begin{equation}
\mathbf{E}_{3} = \bm{\underline{A}}_{3}\cdot\left[e^{-ik_{z,3}z}\bm{\underline{I}}\right]\cdot\mathbf{E}_{0,3}.
\end{equation}
By considering upward and downward propagating waves in each layer one can compute the unknown matrices. The downward propagating wave in layer 2 is a consequence of the transmitted wave from layer 1 and the reflected wave from layer 2 - layer 3 interface
\begin{multline}
e^{-ik_{z,2}d_{12}}\mathbf{E}_{0,2} = e^{-ik_{z,1}d_{12}}\bm{\underline{T}}_{12}\cdot\mathbf{E}_{0,1}\\
+ e^{ik_{z,2}(d_{12}-2d_{23})}\bm{\underline{R}}_{21}\cdot\bm{\underline{R}}_{23}\cdot\mathbf{E}_{0,2},
\end{multline}
which can be solved for $\mathbf{E}_{0,2}$ to give
\begin{equation}
\mathbf{E}_{0,2} = e^{-i(k_{z,1}-k_{z,2})d_{12}}\bm{\underline{M}}^{-1}_{2123}\cdot\bm{\underline{T}}_{12}\cdot\mathbf{E}_{0,1},
\label{E2}
\end{equation}
where
\begin{equation}
\bm{\underline{M}}_{2123} = \bm{\underline{I}} - e^{2ik_{z2}(d_{12}-d_{23})}\bm{\underline{R}}_{21}\cdot\bm{\underline{R}}_{23},
\end{equation}
and the power of $-1$ implies the matrix inverse. The upward propagating wave in layer 1 is a combination of reflected waves from the layer 1 - layer 2 interface and transmitted waves from layer 2
\begin{multline}
e^{-ik_{z,1}d_{12}}\tilde{\bm{\underline{R}}}_{12}\cdot\mathbf{E}_{0,1} = e^{-ik_{z,1}d_{12}}\bm{\underline{R}}_{12}\cdot\mathbf{E}_{0,1}\\
 + e^{ik_{z,2}(d_{12}-2d_{23})}\bm{\underline{T}}_{21}\cdot\bm{\underline{R}}_{23}\cdot\mathbf{E}_{0,2},
\end{multline}
which with the help of Eq.~\eqref{E2} can be used to solve for the generalized reflection coefficient
\begin{equation}
\tilde{\bm{\underline{R}}}_{12} = \bm{\underline{R}}_{12} + e^{-2ik_{z,2}(d_{23} - d_{12})}\bm{\underline{T}}_{21}\cdot\bm{\underline{R}}_{23}\cdot \bm{\underline{M}}^{-1}_{2123}\cdot\bm{\underline{T}}_{12}.
\end{equation}
Adding further layers below layer 3 merely requires one to replace $\bm{\underline{R}}_{23}$ with $\tilde{\bm{\underline{R}}}_{23}$. Thus, one obtains a recursive relation for the reflection coefficient
\begin{widetext}
\begin{equation}
\tilde{\bm{\underline{R}}}_{i,i+1} = \bm{\underline{R}}_{i,i+1} + e^{-2ik_{z,i+1}(d_{i+1,i+2} - d_{i,i+1})}\bm{\underline{T}}_{i+1,i}\cdot\tilde{\bm{\underline{R}}}_{i+1,i+2}\cdot \bm{\underline{M}}^{-1}_{i+1,i,i+1,i+2}\cdot\bm{\underline{T}}_{i,i+1},
\end{equation}
and hence one can generate the reflection coefficient for a medium with any number of layers. For transmission, from Eq. \eqref{E2}, downward going waves in layer 2 are given by
\begin{equation}
e^{-ik_{z,2}d_{12}}\mathbf{E}_{0,2} = e^{-ik_{z,1}d_{12}}\bm{\underline{M}}^{-1}_{2123}\cdot\bm{\underline{T}}_{12}\cdot\mathbf{E}_{0,1},
\end{equation}
and similarly for the following layer
\begin{equation}
e^{-ik_{z,3}d_{23}}\mathbf{E}_{0,3} = e^{-ik_{z,2}d_{23}}\bm{\underline{M}}^{-1}_{3234}\cdot\bm{\underline{T}}_{23}\cdot\mathbf{E}_{0,2}
 = e^{-ik_{z,2}(d_{23}-d_{12})}e^{-ik_{z,1}d_{12}}\bm{\underline{M}}^{-1}_{3234} \cdot\bm{\underline{T}}_{23}\cdot\bm{\underline{M}}^{-1}_{2123} \cdot\bm{\underline{T}}_{12}\cdot\mathbf{E}_{0,1}.
\end{equation}
Thus, one can see that the generalized transmission coefficient reads
\begin{equation}
\tilde{\bm{\underline{T}}}_{1N} = e^{ik_{z,N}(d_{N,N+1}-d_{N-1,N})}\prod^{N}_{i=2}e^{-ik_{z,i}(d_{i,i+1}-d_{i-1,i})}\bm{\underline{M}}^{-1}_{i,i-1,i,i+1}\cdot\bm{\underline{T}}_{i-1,i}.
\end{equation}
\end{widetext}

\section{Embedded Sources}

\begin{figure}[h]
\centering
\includegraphics[width=0.8\linewidth]{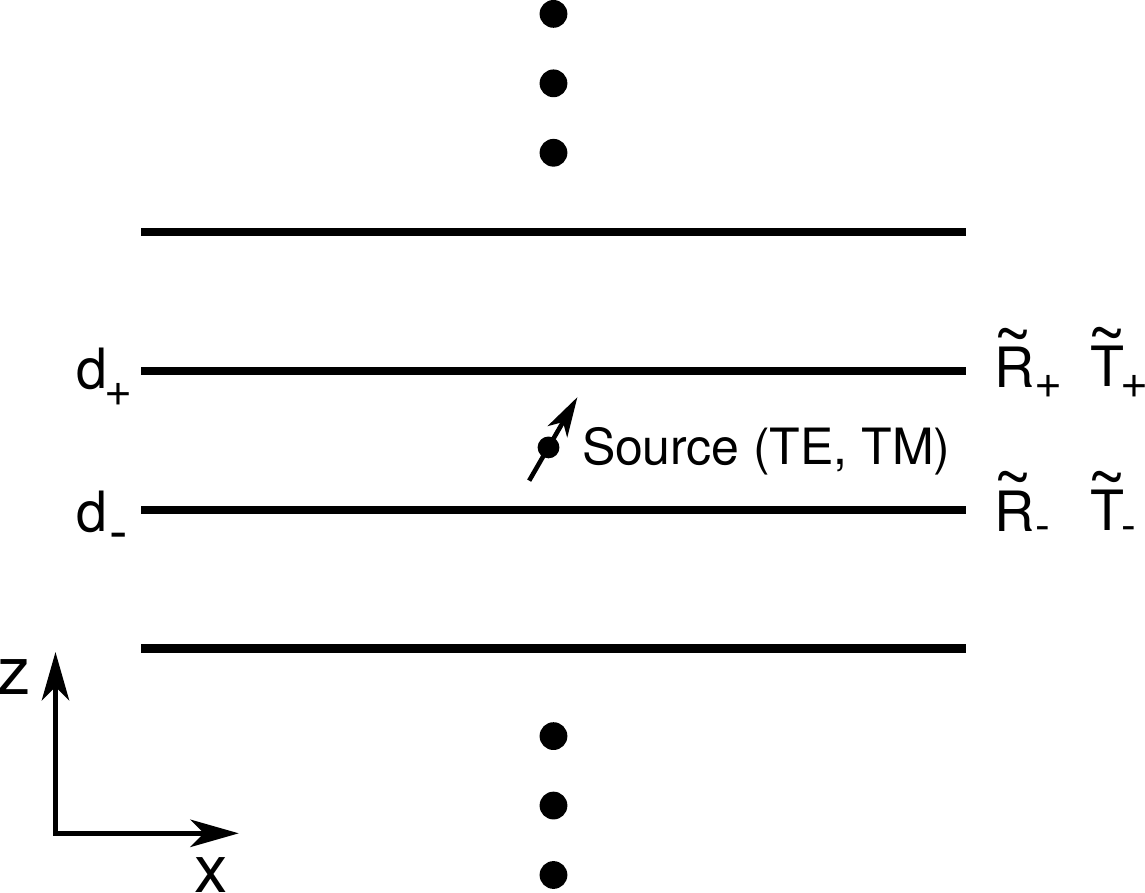}
\caption{Layered media with an embedded source.}
\end{figure}
Consider a source embedded in a layered media. The source produces a unit amplitude electric field with polarization, $\mathbf{E}_{0}$, in the $TE/TM$ basis. The general expression for the field $\mathbf{E}$ in the same layer is given by
\begin{equation}
\mathbf{E} = \mathbf{F}\cdot\mathbf{E}_{0} = \left[e^{ik_{z}|z-z'|}\bm{\underline{I}} + e^{-ik_{z}z}\bm{\underline{B}} + e^{ik_{z}z}\bm{\underline{D}}\right]\cdot\mathbf{E}_{0}.
\end{equation}
The terms with coefficients $\bm{\underline{B}}$ correspond to downward propagating waves and is a result of reflections from the surface at $d_{+}$ and the terms with coefficients $\bm{\underline{D}}$ correspond to upward propagating waves and is a result of reflections from the surface at $d_{-}$. Thus, at the upper interface, we have
\begin{gather}
\bm{\underline{B}}e^{-ik_{z}d_{+}} = \tilde{\bm{\underline{R}}}_{+}\cdot\left[\bm{\underline{I}}e^{ik_{z}|d_{+}-z'|}+\bm{\underline{D}}e^{ik_{z}d_{+}}\right],
\end{gather}
where $\tilde{\bm{\underline{R}}}_{+}$ is the generalized reflection matrix at the $d_{+}$ surface and at the lower interface, we have
\begin{gather}
\bm{\underline{D}}e^{ik_{z}d_{-}} = \tilde{\bm{\underline{R}}}_{-}\cdot\left[\bm{\underline{I}}e^{ik_{z}|d_{-}-z'|}+\bm{\underline{B}}e^{-ik_{z}d_{-}}\right],
\end{gather}
where $\tilde{\bm{\underline{R}}}_{-}$ is the reflection matrix at the $d_{-}$ surface. Solving for $\bm{\underline{B}}$ and $\bm{\underline{D}}$ gives
\begin{multline}
\bm{\underline{B}}e^{-ik_{z}d_{+}} = \tilde{\bm{\underline{M}}}_{+-}\cdot\left[e^{ik_{z}|d_{+}-z'|}\tilde{\bm{\underline{R}}}_{+}\right.\\
\left. +e^{ik_{z}(d_{+}-d_{-})}e^{ik_{z}|d_{-}-z'|}\tilde{\bm{\underline{R}}}_{+}\cdot\tilde{\bm{\underline{R}}}_{-}\right],
\end{multline}
and
\begin{multline}
\bm{\underline{D}}e^{ik_{z}d_{-}} =\tilde{\bm{\underline{M}}}_{-+}\cdot\left[e^{ik_{z}|d_{-}-z'|}\tilde{\bm{\underline{R}}}_{-}\right.\\
\left.  +e^{ik_{z}(d_{+}-d_{-})}e^{ik_{z}|d_{+}-z'|}\tilde{\bm{\underline{R}}}_{-}\cdot\tilde{\bm{\underline{R}}}_{+}\right],
\end{multline}
with the multiple reflection coefficient reading
\begin{equation}
\tilde{\bm{\underline{M}}}_{ij} = \left[\bm{\underline{I}} - e^{2ik_{z}(d_{+}-d_{-})}\tilde{\bm{\underline{R}}}_{i}\cdot\tilde{\bm{\underline{R}}}_{j}\right]^{-1}.
\label{M}
\end{equation}
Again, the power of $-1$ denotes the matrix inverse. Note that, unlike standard magneto-dielectrics, in general $\tilde{\bm{\underline{M}}}_{+-} \neq \tilde{\bm{\underline{M}}}_{-+}$. Substituting the expressions for $\bm{\underline{B}}$ and $\bm{\underline{D}}$ back into the expression for the field, noting that $z,z'>d^{-}$ and $z,z'<d^{+}$and using the definition in Eq. \eqref{M} leads to
\begin{multline}
\mathbf{F}_{z>z'}(z,z') =\\
 e^{ik_{z}z}\tilde{\bm{\underline{M}}}_{-+}\cdot\left[e^{-ik_{z}z'}\bm{\underline{I}} +e^{ik_{z}(z'-2d_{-})}\tilde{\bm{\underline{R}}}_{-}\right]\\
+e^{-ik_{z}z}\tilde{\bm{\underline{M}}}_{+-}\cdot\left[e^{-ik_{z}(z'-2d_{+})}\tilde{\bm{\underline{R}}}_{+}\right.\\
\left. +e^{ik_{z}(z'+2d_{+}-2d_{-})}\tilde{\bm{\underline{R}}}_{+}\cdot\tilde{\bm{\underline{R}}}_{-}\right],
\label{R+}
\end{multline}
for $z>z'$ and
\begin{multline}
\mathbf{F}_{z<z'}(z,z') =\\
e^{-ik_{z}z}\tilde{\bm{\underline{M}}}_{+-}\cdot\left[e^{ik_{z}z'}\bm{\underline{I}} +e^{-ik_{z}(z'-2d_{+})}\tilde{\bm{\underline{R}}}_{+}\right]\\
+e^{ik_{z}z}\tilde{\bm{\underline{M}}}_{-+}\cdot\left[e^{ik_{z}(z'-2d_{-})}\tilde{\bm{\underline{R}}}_{-}\right.\\
\left. +e^{-ik_{z}(z'+2d_{-}-2d_{+})}\tilde{\bm{\underline{R}}}_{-}\cdot\tilde{\bm{\underline{R}}}_{+}\right],
\label{R-}
\end{multline}
for $z<z'$.

One can also find the electric field in a different layer from the source by considering the transmitted fields. The general expression for the field in layer $n$ as a result if a source in layer $m<n$ is
\begin{equation}
\mathbf{E} = \mathbf{F}\cdot\mathbf{E}_{n} = \left[e^{ik_{n,z}z}\bm{\underline{I}} + e^{-ik_{n,z}(z-2d_{n+})}\tilde{\bm{\underline{R}}}_{+}\right]\cdot\mathbf{E}_{n}.
\end{equation}
The upward going field in layer $n$ at the $d_{n-}$ interface can be written as
\begin{align}
\mathbf{E}_{n} &= e^{-ik_{m,z}d_{n-}}\tilde{\bm{\underline{T}}}_{mn}\cdot\mathbf{E}_{m}\nonumber\\
&\qquad + e^{2ik_{m,z}(d_{n+}-d_{n-})}\tilde{\bm{\underline{R}}}_{n-}\cdot\tilde{\bm{\underline{R}}}_{n+}\cdot\mathbf{E}_{n}\nonumber\\
&= e^{-ik_{m,z}d_{n-}}\tilde{\bm{\underline{M}}}_{n-n+}\cdot\tilde{\bm{\underline{T}}}_{mn}\cdot\mathbf{E}_{m},
\end{align}
where the field $\mathbf{E}_{m}$ at the $d_{m+}$ interface is given from Eq. \eqref{R+} by
\begin{multline}
\mathbf{E}_{m} = e^{ik_{m,z}d_{m+}}\tilde{\bm{\underline{M}}}_{m-m+}\cdot\\
\left[e^{-ik_{m,z}z'}\bm{\underline{I}} +e^{ik_{m,z}(z'-2d_{m-})}\tilde{\bm{\underline{R}}}_{m-}\right]\cdot\mathbf{E}_{0}.
\end{multline}
Thus, the $\mathbf{F}$ matrix for the field in layer $n$ is given by
\begin{multline}
\mathbf{F}_{z>z'}(z,z') = \left[e^{ik_{n,z}z}\bm{\underline{I}} + e^{-ik_{n,z}(z-2d_{n+})}\tilde{\bm{\underline{R}}}_{+}\right]\cdot\\
\left[e^{-ik_{m,z}d_{n-}}\tilde{\bm{\underline{M}}}_{n-n+}\cdot\tilde{\bm{\underline{T}}}_{mn}\cdot\tilde{\bm{\underline{M}}}_{m-m+}e^{ik_{m,z}d_{m+}}\right]\cdot \\
\left[e^{-ik_{m,z}z'}\bm{\underline{I}} +e^{ik_{m,z}(z'-2d_{m-})}\tilde{\bm{\underline{R}}}_{m-}\right].
\label{T+}
\end{multline}
Similarly, the general expression for the field in layer $n$ as a result if a source in layer $m>n$ is
\begin{equation}
\mathbf{E} = \mathbf{F}\cdot\mathbf{E}_{n} = \left[e^{-ik_{n,z}z}\bm{\underline{I}} + e^{ik_{n,z}(z-2d_{n-})}\tilde{\bm{\underline{R}}}_{-}\right]\cdot\mathbf{E}_{n}.
\end{equation}
The downward going field in layer $n$ at the $d_{n+}$ interface can be written as
\begin{align}
\mathbf{E}_{n}&= e^{ik_{m,z}d_{n+}}\tilde{\bm{\underline{T}}}_{mn}\cdot\mathbf{E}_{m}\nonumber\\
&\qquad + e^{2ik_{m,z}(d_{n+}-d_{n-})}\tilde{\bm{\underline{R}}}_{n+}\cdot\tilde{\bm{\underline{R}}}_{n-}\cdot\mathbf{E}_{n}\nonumber\\
& = e^{ik_{m,z}d_{n+}}\tilde{\bm{\underline{M}}}_{n+n-}\cdot\tilde{\bm{\underline{T}}}_{mn} \cdot\mathbf{E}_{m},
\end{align}
where the field $\mathbf{E}_{m}$ at the $d_{m+}$ interface is given from Eq. \eqref{R-} by
\begin{multline}
\mathbf{E}_{m} = e^{-ik_{m,z}d_{m+}}\tilde{\bm{\underline{M}}}_{m+m-}\cdot\nonumber\\
\left[e^{ik_{m,z}z'}\bm{\underline{I}} +e^{-ik_{m,z}(z'-2d_{m+})}\tilde{\bm{\underline{R}}}_{m+}\right]\cdot\mathbf{E}_{0}.
\end{multline}
Thus, the $\mathbf{F}$ matrix for the field in layer $n$ is given by
\begin{multline}
\mathbf{F}_{z<z'}(z,z') = \left[e^{-ik_{n,z}z}\bm{\underline{I}} + e^{ik_{n,z}(z-2d_{n-})}\tilde{\bm{\underline{R}}}_{-}\right]\cdot\\ \left[e^{ik_{m,z}d_{n+}}\tilde{\bm{\underline{M}}}_{n+n-}\cdot\tilde{\bm{\underline{T}}}_{mn}\cdot \tilde{\bm{\underline{M}}}_{m+m-}e^{-ik_{m,z}d_{m+}}\right]\cdot\\
\left[e^{ik_{m,z}z'}\bm{\underline{I}} +e^{-ik_{m,z}(z'-2d_{m+})}\tilde{\bm{\underline{R}}}_{m+}\right].
\label{T-}
\end{multline}

\section{The Multilayered Green's Function}

The Green's function is the solution to the wave equation, Eq.~\eqref{EHelmholtz}, for a single frequency point source. For a homogeneous axionic coupling this wave equation reduces to
\begin{equation}
\bm{\nabla}\times\frac{1}{\mu(\mathbf{r},\omega)}\bm{\nabla}\times\mathbf{E}(\mathbf{r},\omega)-\frac{\omega^2}{c^2}\varepsilon(\mathbf{r},\omega)\mathbf{E}(\mathbf{r},\omega) = i\omega\mu_{0}\mathbf{J}(\mathbf{r}),
\label{EHelmholtz2}
\end{equation}
where $\mathbf{J}(\mathbf{r}) = \mathbf{J}_{E}(\mathbf{r},\omega) + \mathbf{J}_{N}(\mathbf{r},\omega)$ is the total current source with both external and noise contributions. Thus, the Green's function is defined by 
\begin{multline}
\bm{\nabla}\times\frac{1}{\mu(\mathbf{r},\omega)}\bm{\nabla}\times\bm{G}(\mathbf{r},\mathbf{r}',\omega)\\
-\frac{\omega^2}{c^2}\varepsilon(\mathbf{r},\omega)\bm{G}(\mathbf{r},\mathbf{r}',\omega) = \delta\left(\mathbf{r} - \mathbf{r}'\right).
\end{multline}
Knowledge of the Green's function allows one to compute the electric field at any point from an arbitrary distribution of current sources via
\begin{equation}
\mathbf{E}(\mathbf{r},\omega) = i\omega\mu_{0}\int d^{3}r'\,\bm{G}(\mathbf{r},\mathbf{r}',\omega)\cdot\mathbf{J}(\mathbf{r}').
\label{EGreen}
\end{equation}

As can be seen from Eq. \eqref{EHelmholtz2}, the wave equation for a homogenous axionic coupling is just the usual wave equation for a traditional magneto-dielectric. Hence the Green's function is identical to the standard magneto-dielectric electric Green's function, which, in its singularity extracted form, reads 
\begin{multline}
\bm{G}(\mathbf{r},\mathbf{r}',\omega) = \frac{i}{8\pi^{2}}\int d^{2}k_{p}\,\mu(\mathbf{r}')\left[\frac{\mathbf{m}^{\ast}(\mathbf{r})\otimes\mathbf{m}(\mathbf{r}')}{k_{z}k^{2}_{p}}\right.\\
\left.+\frac{\mathbf{n}^{\ast}(\mathbf{r})\otimes\mathbf{n}(\mathbf{r}')}{k_{z}k^{2}_{p}}\right]e^{i\mathbf{k}_{p}\cdot(\mathbf{r}_{p}-\mathbf{r}'_{p})}e^{ik_{z}|z-z'|},
\end{multline}
where $\mathbf{m}(\mathbf{r})$ and $\mathbf{n}(\mathbf{r})$ are the dyadic operators
\begin{gather}
\mathbf{m}(\mathbf{r}) = i\bm{\nabla}_{\mathbf{r}}\times\hat{z},\\
\mathbf{n}(\mathbf{r}) = \frac{1}{k}\bm{\nabla}_{\mathbf{r}}\times\bm{\nabla}_{\mathbf{r}}\times\hat{z},
\end{gather}
that generate the solenoidal vector wave functions \cite{chew}, which are equivalent to the polarization vectors in \cite{buhmann}. Here, $\mathbf{k}$ is the wavevector of the wave with $\mathbf{k}_{p} = k_{x}\hat{x}+k_{y}\hat{y}$ and $k_{z} = \sqrt{k^{2}-k_{p}^{2}}$. Similarly, $\mathbf{r}_{p} = r_{x}\hat{x}+r_{y}\hat{y}$. For simplicity we have neglected the source singularity.

The effect of the axionic coupling is only seen when there are inhomogeneities in the material. Adding planar layers is identical to finding generalized reflection coefficients except now we replace the source $\mathbf{E}_{0}$ with a vector containing the dyads. Thus the Green's function for layered TSB-TI's is given by
\begin{multline}
\bm{G}_{z\gtrless z'}(\mathbf{r},\mathbf{r}',\omega) = \frac{i}{8\pi^{2}}\int d^{2}k_{p}\\
\times\mu(\mathbf{r}')\left[\frac{\mathbf{C}(\mathbf{r},\mathbf{r}'):\bm{\underline{F}}_{z\gtrless z'}(z,z')}{k_{z}k^{2}_{p}}\right]e^{i\mathbf{k}_{p}\cdot(\mathbf{r}_{p}-\mathbf{r}'_{p})},
\label{green}
\end{multline}
with
\begin{equation}
\mathbf{C}(\mathbf{r},\mathbf{r}') = \left(\begin{array}{cc}
\mathbf{m}^{\ast}(\mathbf{r})\otimes\mathbf{m}(\mathbf{r}') & \mathbf{n}^{\ast}(\mathbf{r})\otimes\mathbf{m}(\mathbf{r}')\\
\mathbf{m}^{\ast}(\mathbf{r})\otimes\mathbf{n}(\mathbf{r}') &\mathbf{n}^{\ast}(\mathbf{r})\otimes\mathbf{n}(\mathbf{r}')
\end{array}\right),
\end{equation}
and the $:$ operator implying the element-wise Frobenius inner product. For $z$ and $z'$ in the same layer $\underline{\bm{F}}_{z>z'}(z,z')$ is given by Eq. \eqref{R+} and $\underline{\bm{F}}_{z<z'}(z,z')$ by Eq. \eqref{R-}. For $z \in n$ and $z' \in m$ in the different layers $\underline{\bm{F}}_{z>z'}(z,z')$ is given by Eq. \eqref{T+} and $\underline{\bm{F}}_{z<z'}(z,z')$ by Eq. \eqref{T-}. As a consistency check, one can show that the resulting Green's function reduces to that for a traditional magneto-dielectric material when the axion coupling vanishes and that it satisfies the Schwarz reflection principle, which is required for the response to be causal (see Appendix \ref{SRP}).

\section{Dipole Fields Close to a TSB-TI Surface}

As an example of the use of the Green's function we will compute the electric field pattern of a single frequency, dipole point source close to a TSB-TI surface at $z=0$. We take the source to be in the upper layer, $z'>0$. For a field point at $z>0$ there are two contributions, one from direct propagation from the source to the field point, which is given by the free space Green's function $\bm{G}_{0}(\mathbf{r},\mathbf{r}',\omega)$, and one from reflections from the surface, which is given by the reflective part of the Green's function $\bm{R}(\mathbf{r},\mathbf{r}',\omega)$. For $z<0$ the only contribution is from transmission at the surface, which is given by the transmissive part of the Greens function $\bm{T}(\mathbf{r},\mathbf{r}',\omega)$. Thus, we can split the Green's function into 3 parts
\begin{equation}
\bm{G}(\mathbf{r},\mathbf{r}',\omega) = 
\left\{\begin{array}{c}
\bm{G}_{0}(\mathbf{r},\mathbf{r}',\omega) + \bm{R}(\mathbf{r},\mathbf{r}',\omega) \qquad z>0,\\
\bm{T}(\mathbf{r},\mathbf{r}',\omega) \qquad z<0,
\end{array}\right.
\end{equation}
each of which can be computed separately. Each part can be found by expanding the definition of the Green's function given in Eq. \eqref{green}. These expressions are given in Appendix \ref{App1}.

\subsection{z-Orientated-Dipole}

\begin{figure*}[t]
\centering
\includegraphics[width=0.9\linewidth]{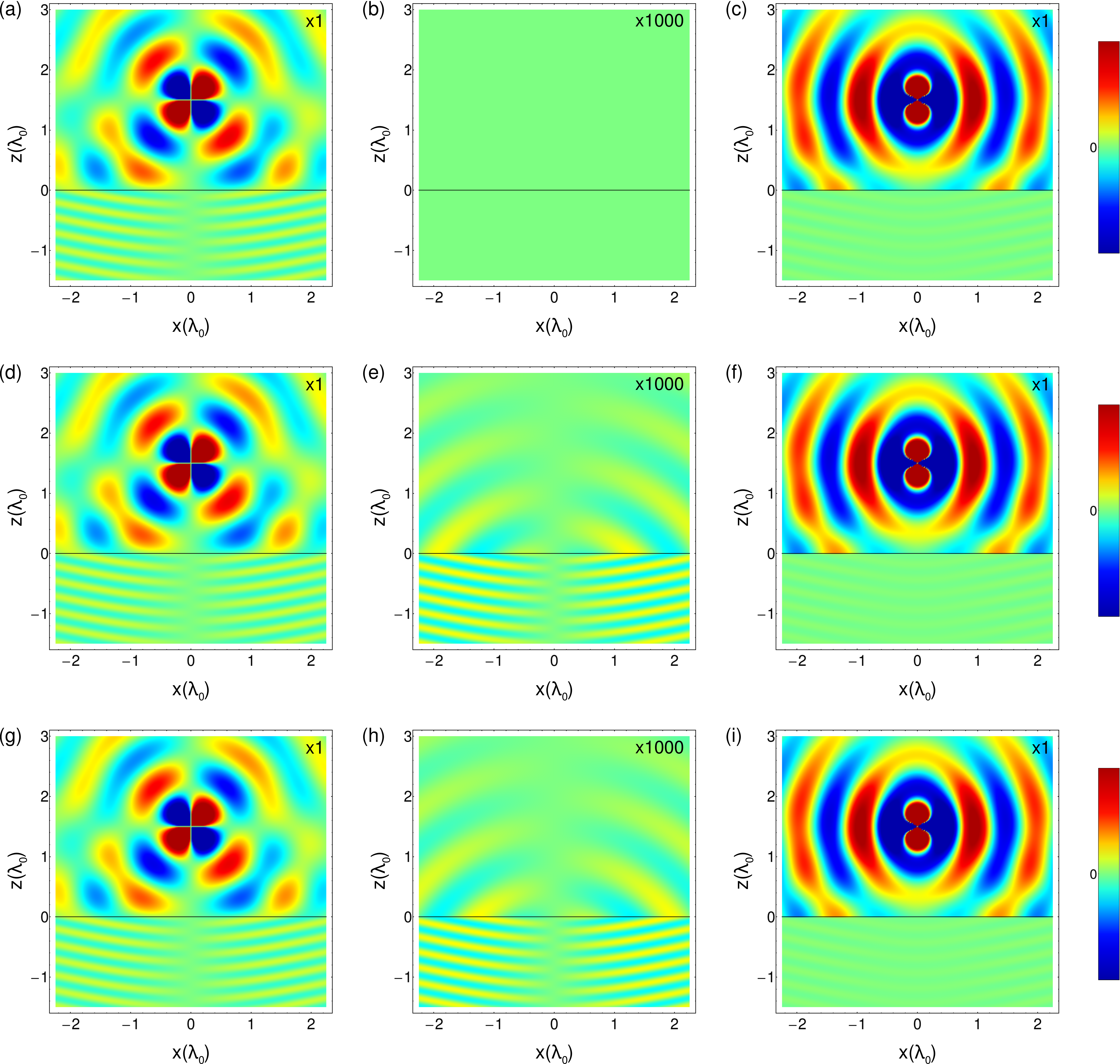}
\caption{(Color online) The field pattern (real part) in the $x-z$ plane for a single frequency, $\hat{z}$ orientated, point dipole close to a TSB-TI interface. Here, the upper layer is the vacuum ($\mu = 1$, $\varepsilon = 1$ and $\Theta = 0$) and lower layer is a medium with $\varepsilon = 16$ and $\mu = 1$. (a), (b) and (c) are the $x$, $y$ and $z$ components, respectively, when the axion coupling in the medium is $\Theta = 0$. (d), (e) and (f) are the $x$, $y$ and $z$ components, respectively, when the axion coupling in the medium is $\Theta = \pi$. (g), (h) and (i) are the $x$, $y$ and $z$ components, respectively, when the axion coupling in the medium is $\Theta = -\pi$. All distance are in terms of the vacuum wavelength and all field strength units are arbitrary. Note that, for clarity, the amplitude of the $y$ component of the field has been scaled by $\times 1000$ compared to the $x$ and $z$ components.}
\label{zdipole}
\end{figure*}

First we will consider a single frequency, dipole point source, orientated in the $z$ direction, placed close to a material surface. This source can be represented by a current density of the form
\begin{equation}
\mathbf{J}(\mathbf{r}') = -i\omega d(\omega)\delta(\mathbf{r}')\hat{z},
\end{equation} 
where $d(\omega)$ is the dipole strength. The source is placed in the upper layer, which is taken to be the vacuum ($\varepsilon = 1$ and $\mu = 1$), at $1.5\,\lambda_{0}$ above a surface, where $\lambda_{0}$ is the vacuum wavelength. The material parameters for the surface are $\varepsilon = 16$ and $\mu = 1$, which are comparable to those of $\mathrm{Bi_{2}Se_{3}}$ \cite{bi2se3}. Substituting the current source into the expression for the electric field in Eq. \eqref{EGreen} shows that the relevant components of the Green's function are the $G^{iz}(\mathbf{r},\mathbf{r}',\omega)$, where $i = x,y,z$ depending on the desired field component at $\mathbf{r}$. The expression for the Green's function components can be simplified by converting to polar coordinates, after which the angular integral can be performed analytically. The resulting Hankel transform integral, however, must be computed numerically (the relevant integrals can be found in Appendix \ref{App2}). 

The field patterns for this configuration are shown in Figure \ref{zdipole}. Figures \ref{zdipole} (a), (b) and (c) show the, $x$, $y$ and $z$ components, respectively, for the real part of the electric field (equivalent to the time dependent field at $t=0$) in the $x-z$ plane for $\Theta = 0$ - the case of a conventional magneto-dielectric. In this case the mixing coefficients vanish and hence one sees $y$-component of the field is zero. The field patterns for the $x$ and $z$ components are those that one would expect from a point dipole. Figures \ref{zdipole} (d), (e) and (f) show the, $x$, $y$ and $z$ components, respectively, for the real part of the electric field in the $x-z$ plane for the case of a TSB-TI with, $\Theta = \pi$. In this case the mixing coefficients are non-zero. The axion coupling causes a rotation of the polarization of the field, generating a non-zero $y$-component at the interface. Figures \ref{zdipole} (g), (h) and (i) show the, $x$, $y$ and $z$ components, respectively, for the real part of the electric field in the $x-z$ plane for the case of a TSB-TI with, $\Theta = -\pi$. This case similar to that of a a TSB-TI with $\Theta = \pi$, except that the interface causes the field polarization to be rotated in the opposite direction. Thus, the $y$-component is the opposite of that in Figure \ref{zdipole} (e). Note that the discontinuity at the interface in Figures \ref{zdipole} (c), (f) and (i) is expected since this is the longitudinal component of the electric field, which, unlike the transverse components, is not continuous at the boundary. In fact, as the difference in permittivity at the interface is $16$, one would expect an order of magnitude jump in the longitudinal component, which is observed.

\subsection{x-Orientated-Dipole}

\begin{figure*}[t]
\centering
\includegraphics[width=0.9\linewidth]{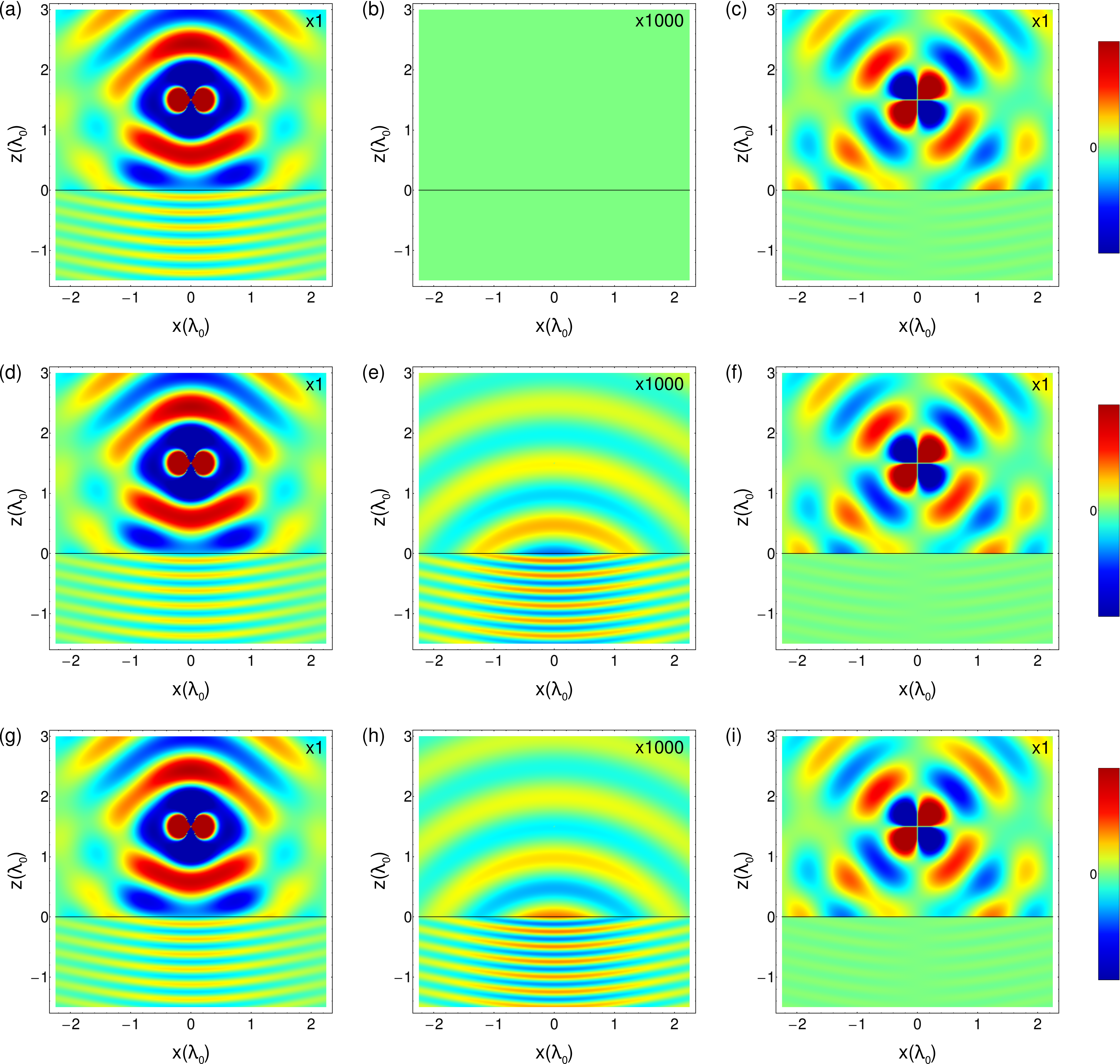}
\caption{(Color online) The field pattern (real part) in the $x-z$ plane for a single frequency, $\hat{x}$ orientated, point dipole close to a TSB-TI interface. Here, the upper layer is the vacuum ($\mu = 1$, $\varepsilon = 1$ and $\Theta = 0$) and lower layer is a medium with $\varepsilon = 16$ and $\mu = 1$. (a), (b) and (c) are the $x$, $y$ and $z$ components, respectively, when the axion coupling in the medium is $\Theta = 0$. (d), (e) and (f) are the $x$, $y$ and $z$ components, respectively, when the axion coupling in the medium is $\Theta = \pi$. (g), (h) and (i) are the $x$, $y$ and $z$ components, respectively, when the axion coupling in the medium is $\Theta = -\pi$. All distance are in terms of the vacuum wavelength and all field strength units are in arbitrary. Note that, for clarity, the amplitude of the $y$ component of the field has been scaled by $\times 1000$ compared to the $x$ and $z$ components.}
\label{xdipole}
\end{figure*}

Next we will consider a dipole source orientated in the $x$ direction, which is given by a current density of the form
\begin{equation}
\mathbf{J}(\mathbf{r}') = -i\omega d(\omega)\delta(\mathbf{r}')\hat{x},
\end{equation} 
where, again, $d(\omega)$ is the dipole strength. We consider a similar geometry as before with the source was placed in the upper layer (vacuum) at $1.5\,\lambda_{0}$ above a surface. The material parameters for the surface are, again, $\varepsilon = 16$ and $\mu = 1$. Substitution of the source current into the expression for the electric field in Eq. \eqref{EGreen} shows that the relevant components of the Green's function, in this case, are the $G^{ix}(\mathbf{r},\mathbf{r}',\omega)$, where $i \in x,y,z$ depending on the desired field component at $\mathbf{r}$. Again, by converting to polar coordinates and computing the angular integral one arrives at a Hankel transform integral that must be computed numerically. The relevant integrals can be found in Appendix \ref{App3}.

The field patterns for this configuration are shown in Figure \ref{xdipole}. Figures \ref{xdipole} (a), (b) and (c) show the, $x$, $y$ and $z$ components, respectively, for the real ($t=0$) part of the electric field in the $x-z$ plane for a conventional magneto-dielectric ($\Theta = 0$). Here, as with the $z$-orientated dipole, the mixing coefficients vanish and hence one sees no $y$-component to the field. Figures \ref{xdipole} (d), (e) and (f) show the, $x$, $y$ and $z$ components, respectively, for the real part of the electric field in the $x-z$ plane for the case of a TSB-TI with, $\Theta = \pi$. One, again, sees the generation of a non-zero $y$-component owing to the effects of the interface. Figures \ref{xdipole} (g), (h) and (i) show the, $x$, $y$ and $z$ components, respectively, for the real part of the electric field in the $x-z$ plane for the case of a TSB-TI with, $\Theta = -\pi$. As before we see the inversion of the $y$-component compared to that in Figure \ref{xdipole} (e). Again, the discontinuity in Figures \ref{xdipole} (c), (f) and (i) is expected since this is the longitudinal component of the electric field.

\section{Summary}

We have constructed the Green's function of a layered TSB-TI and used it to study the field pattern of a single frequency point dipole close to the surface of a topological insulator. Reflection and transmission from a TSB-TI surface leads to mixing of $TE$ and $TM$ polarization components and hence a rotation in the overall polarization of the incident light. This effect has the potential to be the basis for a number of novel optical and quantum optical effects, for whose study the Green's function will be useful. Owing to the ubiquitous nature of the Green's function in both classical and quantum electromagnetism, it is hoped that the closed form expressions for this function will prove to be beneficial to a great many fields.

\section{Acknowledgements}

This work was supported by the DFG (grants BU 1803/3-1 and GRK 2079/1). SYB is grateful for support by the Freiburg Institute for Advanced Studies. 

\appendix

\section{Schwarz Reflection Principle}
\label{SRP}

The Schwarz reflection principle states that
\begin{equation}
\bm{G}^{\ast}(\mathbf{r},\mathbf{r}',\omega) = \bm{G}(\mathbf{r},\mathbf{r}',-\omega^{\ast}).
\end{equation}
One can show that the same principle holds for axionic materials. Given the Helmholtz equation in Eq. \eqref{EHelmholtz}, the definition of the Green's function reads
\begin{align}
&\bm{\nabla}\times\frac{1}{\mu(\mathbf{r},\omega)}\bm{\nabla}\times\bm{G}(\mathbf{r},\mathbf{r}',\omega)\nonumber\\
&\qquad -i\frac{\omega}{c}\frac{\alpha}{\pi}\left[\bm{\nabla}\Theta(\mathbf{r},\omega)\times\bm{G}(\mathbf{r},\mathbf{r}',\omega)\right]\nonumber\\
&\qquad\qquad -\frac{\omega^{2}}{c^2}\varepsilon(\mathbf{r},\omega)\bm{G}(\mathbf{r},\mathbf{r}',\omega) = \delta(\mathbf{r}-\mathbf{r}').
\label{SRPeq1}
\end{align}
Setting $\omega \rightarrow -\omega^{\ast}$ gives
\begin{align}
&\bm{\nabla}\times\frac{1}{\mu(\mathbf{r},-\omega^{\ast})}\bm{\nabla}\times\bm{G}(\mathbf{r},\mathbf{r}',-\omega^{\ast})\nonumber\\
&\quad +i\frac{\omega^{\ast}}{c}\frac{\alpha}{\pi}\left[\bm{\nabla}\Theta(\mathbf{r},-\omega^{\ast})\times\bm{G}(\mathbf{r},\mathbf{r}',-\omega^{\ast})\right]\nonumber\\
&\qquad -\frac{\left(\omega^{\ast}\right)^{2}}{c^2}\varepsilon(\mathbf{r},-\omega^{\ast})\bm{G}(\mathbf{r},\mathbf{r}',-\omega^{\ast}) = \delta(\mathbf{r}-\mathbf{r}').
\end{align}
As the permittivity, $\varepsilon(\mathbf{r},\omega)$, and permeability, $\mu(\mathbf{r},\omega)$, via causality arguments, also obey the Schwarz reflection principle and the axion coupling, $\Theta(\mathbf{r},\omega)$, is real, we have
\begin{align}
&\bm{\nabla}\times\frac{1}{\mu^{\ast}(\mathbf{r},\omega)}\bm{\nabla}\times\bm{G}(\mathbf{r},\mathbf{r}',-\omega^{\ast})\nonumber\\
&\quad +i\frac{\omega^{\ast}}{c}\frac{\alpha}{\pi}\left[\bm{\nabla}\Theta(\mathbf{r},\omega)\times\bm{G}(\mathbf{r},\mathbf{r}',-\omega^{\ast})\right]\nonumber\\
&\qquad -\frac{\left(\omega^{\ast}\right)^{2}}{c^2}\varepsilon^{\ast}(\mathbf{r},\omega)\bm{G}(\mathbf{r},\mathbf{r}',-\omega^{\ast}) = \delta(\mathbf{r}-\mathbf{r}').
\label{SRPeq2}
\end{align}
By comparing Eq. \eqref{SRPeq2} with the complex conjugate of Eq. \eqref{SRPeq1}, one can see that the Schwarz reflection principle holds for axionic materials.

For purely imaginary frequencies one has
\begin{equation}
\bm{G}^{\ast}(\mathbf{r},\mathbf{r}', i\xi) = \bm{G}(\mathbf{r},\mathbf{r}',-\left(i\xi\right)^{\ast}) = \bm{G}(\mathbf{r},\mathbf{r}', i\xi),
\end{equation}
where $\xi$ is real. Hence at imaginary frequencies the Green's function is real. In addition one has,
\begin{equation}
k_{z} = \sqrt{\frac{(i\xi)^2}{c^2} - k_{p}^{2}} = i\sqrt{\frac{\xi^{2}}{c^{2}}+k_{p}^{2}} = i\kappa_{z},
\end{equation}
where $\kappa_{z}$ is real. Substituting this into the expression for the Green's function in Appendix \ref{App1}, converting to angular coordinates, performing the angular integration and noting that the resulting Hankel transform is real one can see that the components of the planar half space Green's function obey the Schwarz reflection principle.

\section{Greens Function for a Planar Half Space}
\label{App1}

The half space Green's function has a single interface at $z=0$, hence $d_{\pm} = 0$. Hence, the multiple reflection coefficients $\tilde{\bm{\underline{M}}}_{ij} = \bm{\underline{I}}$. For a source point in the upper layer, the upper reflection and transmission coefficients vanish, $\tilde{\bm{\underline{R}}}_{+} = \tilde{\bm{\underline{T}}}_{+} = 0$, and the lower reflection and transmission coefficients becomes that for a single interface, $\tilde{\bm{\underline{R}}}_{-} = \bm{\underline{R}}$ and $\tilde{\bm{\underline{T}}}_{-} = \bm{\underline{T}}$. Similarly, for a source point in the lower layer, the lower reflection and transmission coefficients vanish, $\tilde{\bm{\underline{R}}}_{-} = \tilde{\bm{\underline{T}}}_{-} = 0$, and the upper reflection coefficient becomes that for a single interface, $\tilde{\bm{\underline{R}}}_{+} = \bm{\underline{R}}$ and  $\tilde{\bm{\underline{T}}}_{+} = \bm{\underline{T}}$. Putting this together, expanding the dyads, and ignoring the extracted singularity, leads to expressions for the reflective, transmissive and free space parts of the Green's function for a source and field point in an arbitrary layer.

The free space part of the Green's function is proportional to $\bm{\underline{I}}$ and is given by the well known expression
\begin{equation}
\bm{G}_{0}(\mathbf{r},\mathbf{r}',\omega)=\mu(\omega)\left[\frac{1}{k^2}\bm{\nabla}\otimes\bm{\nabla}+\bm{\underline{I}}\right]\frac{e^{ikR}}{4\pi R},
\end{equation}
which, on evaluating the derivatives, becomes
\begin{multline}
\bm{G}_{0}(\mathbf{r},\mathbf{r}',\omega) = \mu(\omega)\frac{e^{ikR}}{4\pi R}\left[\left(1+\frac{ikR-1}{k^2R^2}\right)\bm{\underline{I}}\right.\\
\left.+\frac{3-3ikR-k^2R^2}{k^2R^2}\frac{\mathbf{R}\otimes\mathbf{R}}{R^2}\right],
\label{FreeGreen}
\end{multline}
with $\mathbf{R} = \mathbf{r}-\mathbf{r}'$.

The reflective part of the Green's function can be written as
\begin{equation}
\bm{G}(\mathbf{r},\mathbf{r}') = \int\frac{d^{2}k_{p}}{(2\pi)^{2}}\,e^{i\mathbf{k}_{p}\cdot(\mathbf{r}_{p}-\mathbf{r}'_{p})}R^{ij}(\mathbf{k}_{p},k_{z},z,z'),
\end{equation}
with
\begin{align}
&R^{xx}(\mathbf{k}_{p},k_{z},z,z') =\frac{i\mu(\omega)}{2k_{z}}e^{ik_{z}(|z|+|z'|)}\nonumber\\
&\qquad\times\left[\frac{k_{y}^{2}}{k^{2}_{p}}R_{TE,TE}-\frac{k_{x}^{2}k_{z}^{2}}{k^{2}_{p}k^{2}}R_{TM,TM}\right.\nonumber\\
&\qquad\left. +\mathrm{sgn}\left(z'\right)\frac{k_{z}}{k^{2}_{p}k}\left(k_{x}k_{y}R_{TE,TM}-k_{x}k_{y}R_{TM,TE}\right)\right],
\end{align}
\begin{align}
&R^{yy}(\mathbf{k}_{p},k_{z},z,z') =\frac{i\mu(\omega)}{2k_{z}}e^{ik_{z}(|z|+|z'|)}\nonumber\\
&\qquad\times\left[\frac{k_{x}^{2}}{k^{2}_{p}}R_{TE,TE}-\frac{k_{y}^{2}k_{z}^{2}}{k^{2}_{p}k^{2}}R_{TM,TM}\right.\nonumber\\
&\qquad\left. -\mathrm{sgn}\left(z'\right)\frac{k_{z}}{k^{2}_{p}k}\left(k_{x}k_{y}R_{TE,TM}-k_{x}k_{y}R_{TM,TE}\right)\right],
\end{align}
\begin{align}
&R^{xy}(\mathbf{k}_{p},k_{z},z,z') =\frac{i\mu(\omega)}{2k_{z}}e^{ik_{z}(|z|+|z'|)}\nonumber\\
&\qquad\times\left[-\frac{k_{x}k_{y}}{k^{2}_{p}}R_{TE,TE}-\frac{k_{x}k_{y}k_{z}^{2}}{k^{2}_{p}k^{2}}R_{TM,TM}\right.\nonumber\\
&\qquad\left. +\mathrm{sgn}\left(z'\right)\frac{k_{z}}{k^{2}_{p}k}\left(k_{y}^{2}R_{TE,TM}+k_{x}^{2}R_{TM,TE}\right)\right],
\end{align}
\begin{align}
&R^{yx}(\mathbf{k}_{p},k_{z},z,z') =\frac{i\mu(\omega)}{2k_{z}}e^{ik_{z}(|z|+|z'|)}\nonumber\\
&\qquad\times\left[-\frac{k_{x}k_{y}}{k^{2}_{p}}R_{TE,TE}-\frac{k_{x}k_{y}k_{z}^{2}}{k^{2}_{p}k^{2}}R_{TM,TM}\right.\nonumber\\
&\qquad\left. -\mathrm{sgn}\left(z'\right)\frac{k_{z}}{k^{2}_{p}k}\left(k_{x}^{2}R_{TE,TM}+k_{y}^{2}R_{TM,TE}\right)\right],
\end{align}
\begin{align}
&R^{xz}(\mathbf{k}_{p},k_{z},z,z') =\frac{i\mu(\omega)}{2k_{z}}e^{ik_{z}(|z|+|z'|)}\nonumber\\
&\qquad\times\left[-\mathrm{sgn}\left(z'\right)\frac{k_{x}k_{z}}{k^{2}}R_{TM,TM}+\frac{k_{y}}{k}R_{TE,TM}\right],
\end{align}
\begin{align}
&R^{zx}(\mathbf{k}_{p},k_{z},z,z') =\frac{i\mu(\omega)}{2k_{z}}e^{ik_{z}(|z|+|z'|)}\nonumber\\
&\qquad\times\left[\mathrm{sgn}\left(z'\right)\frac{k_{x}k_{z}}{k^{2}}R_{TM,TM}+\frac{k_{y}}{k}R_{TM,TE}\right],
\end{align}
\begin{align}
&R^{yz}(\mathbf{k}_{p},k_{z},z,z') =\frac{i\mu(\omega)}{2k_{z}}e^{ik_{z}(|z|+|z'|)}\nonumber\\
&\qquad\times\left[-\mathrm{sgn}\left(z'\right)\frac{k_{y}k_{z}}{k^{2}}R_{TM,TM}-\frac{k_{x}}{k}R_{TE,TM}\right],
\end{align}
\begin{align}
&R^{zy}(\mathbf{k}_{p},k_{z},z,z') =\frac{i\mu(\omega)}{2k_{z}}e^{ik_{z}(|z|+|z'|)}\nonumber\\
&\qquad\times\left[\mathrm{sgn}\left(z'\right)\frac{k_{y}k_{z}}{k^{2}}R_{TM,TM}-\frac{k_{x}}{k}R_{TM,TE}\right],
\end{align}
\begin{equation}
R^{zz}(\mathbf{k}_{p},k_{z},z,z') =\frac{i\mu(\omega)}{2k_{z}}e^{ik_{z}(|z|+|z'|)}\left[\frac{k^{2}_{p}}{k^{2}}R_{TM,TM}\right],
\end{equation}
where $\mu(\omega)$ is the permeability in the layer. 

Similarly, the transmissive part of the Green's function can be written as
\begin{equation}
\bm{G}(\mathbf{r},\mathbf{r}') = \int\frac{d^{2}k_{p}}{(2\pi)^{2}}\,e^{i\mathbf{k}_{p}\cdot(\mathbf{r}_{p}-\mathbf{r}'_{p})}T^{ij}(\mathbf{k}_{p},k_{z},z,z'),
\end{equation}
with
\begin{align}
&T^{xx}(\mathbf{k}_{p},k_{z},z,z') =\frac{i\mu'(\omega)}{2k_{z'}}e^{ik_{z}|z|+ik_{z'}|z'|}\nonumber\\
&\qquad\times\left[\frac{k_{y}^{2}}{k^{2}_{p}}T_{TE,TE}+\frac{k_{x}^{2}k_{z}k_{z'}}{k^{2}_{p}kk'}T_{TM,TM}\right.\nonumber\\
&\qquad\left.+\mathrm{sgn}\left(z'\right)\frac{1}{k^{2}_{p}}\left(k_{x}k_{y}\frac{k_{z'}}{k'}T_{TE,TM}+k_{x}k_{y}\frac{k_{z}}{k}T_{TM,TE}\right)\right],
\end{align}
\begin{align}
&T^{yy}(\mathbf{k}_{p},k_{z},z,z') =\frac{i\mu'(\omega)}{2k_{z'}}e^{ik_{z}|z|+ik_{z'}|z'|}\nonumber\\
&\qquad\times\left[\frac{k_{x}^{2}}{k^{2}_{p}}T_{TE,TE}+\frac{k_{y}^{2}k_{z}k_{z'}}{k^{2}_{p}kk'}T_{TM,TM}\right.\nonumber\\
&\qquad\left.-\mathrm{sgn}\left(z'\right)\frac{1}{k^{2}_{p}}\left(k_{x}k_{y}\frac{k_{z'}}{k'}T_{TE,TM}+k_{x}k_{y}\frac{k_{z}}{k}T_{TM,TE}\right)\right],
\end{align}
\begin{align}
&T^{xy}(\mathbf{k}_{p},k_{z},z,z') =\frac{i\mu'(\omega)}{2k_{z'}}e^{ik_{z}|z|+ik_{z'}|z'|}\nonumber\\
&\qquad\times\left[-\frac{k_{x}k_{y}}{k^{2}_{p}}T_{TE,TE}+\frac{k_{x}k_{y}k_{z}k_{z'}}{k^{2}_{p}kk'}T_{TM,TM}\right.\nonumber\\
&\qquad\left.+\mathrm{sgn}\left(z'\right)\frac{1}{k^{2}_{p}}\left(k_{y}^{2}\frac{k_{z'}}{k'}T_{TE,TM}-k_{x}^{2}\frac{k_{z}}{k}T_{TM,TE}\right)\right],
\end{align}
\begin{align}
&T^{yx}(\mathbf{k}_{p},k_{z},z,z') =\frac{i\mu'(\omega)}{2k_{z'}}e^{ik_{z}|z|+ik_{z'}|z'|}\nonumber\\
&\qquad\times\left[-\frac{k_{x}k_{y}}{k^{2}_{p}}T_{TE,TE}+\frac{k_{x}k_{y}k_{z}k_{z'}}{k^{2}_{p}kk'}T_{TM,TM}\right.\nonumber\\
&\qquad\left.-\mathrm{sgn}\left(z'\right)\frac{1}{k^{2}_{p}}\left(k_{x}^{2}\frac{k_{z'}}{k'}T_{TE,TM}-k_{y}^{2}\frac{k_{z}}{k}T_{TM,TE}\right)\right],
\end{align}
\begin{align}
&T^{xz}(\mathbf{k}_{p},k_{z},z,z') =\frac{i\mu'(\omega)}{2k_{z'}}e^{ik_{z}|z|+ik_{z'}|z'|}\nonumber\\
&\qquad\times\left[\mathrm{sgn}\left(z'\right)\frac{k_{x}k_{z}}{kk'}T_{TM,TM}+\frac{k_{y}}{k'}T_{TE,TM}\right],
\end{align}
\begin{align}
&T^{zx}(\mathbf{k}_{p},k_{z},z,z') =\frac{i\mu'(\omega)}{2k_{z'}}e^{ik_{z}|z|+ik_{z'}|z'|}\nonumber\\
&\qquad\times\left[\mathrm{sgn}\left(z'\right)\frac{k_{x}k_{z'}}{kk'}T_{TM,TM}+\frac{k_{y}}{k}T_{TM,TE}\right],
\end{align}
\begin{align}
&T^{yz}(\mathbf{k}_{p},k_{z},z,z') = \frac{i\mu'(\omega)}{2k_{z'}}e^{ik_{z}|z|+ik_{z'}|z'|}\nonumber\\
&\qquad\times\left[\mathrm{sgn}\left(z'\right)\frac{k_{y}k_{z}}{kk'}T_{TM,TM}-\frac{k_{x}}{k'}T_{TE,TM}\right],
\end{align}
\begin{align}
&T^{zy}(\mathbf{k}_{p},k_{z},z,z') =\frac{i\mu'(\omega)}{2k_{z'}}e^{ik_{z}|z|+ik_{z'}|z'|}\nonumber\\
&\qquad\times\left[\mathrm{sgn}\left(z'\right)\frac{k_{y}k_{z'}}{kk'}T_{TM,TM}-\frac{k_{x}}{k}T_{TM,TE}\right],
\end{align}
\begin{equation}
T^{zz}(\mathbf{k}_{p},k_{z},z,z') =\frac{i\mu'(\omega)}{2k_{z'}}e^{ik_{z}|z|+ik_{z'}|z'|}\left[\frac{k^{2}_{p}}{kk'}T_{TM,TM}\right],
\end{equation}
where $\mu'(\omega)$ is the permeability in the source layer. Note that if the cross-reflection, $R_{TE,TM}$ and $R_{TM,TE}$ and cross-transmission, $T_{TM,TE}$ and $T_{TM,TE}$, coefficients vanish one recovers the Green's function for a conventional magneto-dielectric \cite{dung}.

\section{Green's Function Components for an z-Orientated Dipole}
\label{App2}

The field from a z-orientated dipole close to a TSB-TI can be found from the planar half space Green's function in Appendix \ref{App1}. The relevant components can be simplified by converting to polar coordinate, after which the angular integral can be computed analytically. The resulting Hankel transforms, which must be evaluated numerically, are
\begin{multline}
R^{xz}(\mathbf{r},\mathbf{r}') =\frac{1}{4\pi}\int dk_{p}\, e^{ik_{z}(|z|+|z'|)}\\
\times J_{1}\left(k_{p}R_{p}\right)\frac{k^{2}_{p}}{k^{2}}R_{TM,TM},
\end{multline}
\begin{multline}
R^{yz}(\mathbf{r},\mathbf{r}') =\frac{1}{4\pi}\int dk_{p}\,e^{ik_{z}(|z|+|z'|)}\\
\times J_{1}\left(k_{p}R_{p}\right)\frac{k^{2}_{p}}{kk_{z}}R_{TE,TM},
\end{multline}
\begin{multline}
R^{zz}(\mathbf{r},\mathbf{r}') =\frac{i}{4\pi}\int dk_{p}\,e^{ik_{z}(|z|+|z'|)}\\
\times J_{0}\left(k_{p}R_{p}\right)\frac{k^{3}_{p}}{k^{2}k_{z}}R_{TM,TM},
\end{multline}
for the reflective part and 
\begin{multline}
T^{xz}(\mathbf{r},\mathbf{r}') =-\frac{1}{4\pi}\int dk_{p}\,e^{ik_{z}|z|+ik_{z'}|z'|}\\
\times J_{1}\left(k_{p}R_{p}\right)\frac{k_{p}^{2}k_{z}}{kk'k_{z'}}T_{TM,TM},
\end{multline}
\begin{multline}
T^{yz}(\mathbf{r},\mathbf{r}') =\frac{1}{4\pi}\int dk_{p}\,e^{ik_{z}|z|+ik_{z'}|z'|}\\
\times J_{1}\left(k_{p}R_{p}\right)\frac{k_{p}^{2}}{k'k_{z'}}T_{TE,TM},
\end{multline}
\begin{multline}
T^{zz}(\mathbf{r},\mathbf{r}') =\frac{i}{4\pi}\int dk_{p}\,e^{ik_{z}|z|+ik_{z'}|z'|}\\
\times J_{0}\left(k_{p}R_{p}\right)\frac{k^{3}_{p}}{kk'k_{z'}}T_{TM,TM},
\end{multline}
for the transmissive part, where $\mathbf{R}_{p} = \mathbf{r}_{p}-\mathbf{r}_{p}'$.

\section{Green's Function Components for an x-Orientated Dipole}
\label{App3}

The field from a x-orientated dipole close to a TSB-TI can, again, be found from the planar half space Green's function in Appendix \ref{App1}. Converting to polar coordinate and evaluating the angular integral leads to
\begin{multline}
R^{xx}(\mathbf{r},\mathbf{r}') =\frac{i}{8\pi}\int dk_{p}\,e^{ik_{z}(|z|+|z'|)}\\
\times\left\{\left[J_{0}\left(k_{p}R_{p}\right)+2J_{2}\left(k_{p}R_{p}\right)\right]\frac{k_{p}}{k_{z}}R_{TE,TE}\right.\\
\left.-\left[J_{0}\left(k_{p}R_{p}\right)-2J_{2}\left(k_{p}R_{p}\right)\right]\frac{k_{p}k_{z}}{k^{2}}R_{TM,TM}\right\},
\end{multline}
\begin{multline}
R^{yx}(\mathbf{r},\mathbf{r}') = -\frac{i}{8\pi}\int dk_{p}\,e^{ik_{z}(|z|+|z'|)}\\
\times\left\{\left[J_{0}\left(k_{p}R_{p}\right)-2J_{2}\left(k_{p}R_{p}\right)\right]\frac{k_{p}}{k}R_{TE,TM}\right.\\
\left.+\left[J_{0}\left(k_{p}R_{p}\right)+2J_{2}\left(k_{p}R_{p}\right)\right]\frac{k_{p}}{k}R_{TM,TE}\right\},
\end{multline}
\begin{multline}
R^{zx}(\mathbf{r},\mathbf{r}') =-\frac{1}{4\pi}\int dk_{p}\,e^{ik_{z}(|z|+|z'|)}\\
\times\left\{J_{1}\left(k_{p}R_{p}\right)\frac{k_{p}^{2}}{k^{2}}R_{TM,TM}\right\},
\end{multline}
for the reflective part and
\begin{multline}
T^{xx}(\mathbf{r},\mathbf{r}') =\frac{i}{8\pi}\int dk_{p}\,e^{ik_{z}|z|+ik_{z'}|z'|}\\
\times\left\{\left[J_{0}\left(k_{p}R_{p}\right)+2J_{2}\left(k_{p}R_{p}\right)\right]\frac{k_{p}}{k_{z'}}T_{TE,TE}\right.\\
\left.+\left[J_{0}\left(k_{p}R_{p}\right)-2 J_{2}\left(k_{p}R_{p}\right)\right]\frac{k_{p}k_{z}}{kk'}T_{TM,TM}\right\},
\end{multline}
\begin{multline}
T^{yx}(\mathbf{r},\mathbf{r}') =-\frac{i}{8\pi}\int dk_{p}\,e^{ik_{z}|z|+ik_{z'}|z'|}\\
\times\left\{\left[J_{0}\left(k_{p}R_{p}\right)-2J_{2}\left(k_{p}R_{p}\right)\right]\frac{k_{p}}{k'}T_{TE,TM}\right.\\
\left. -\left[J_{0}\left(k_{p}R_{p}\right)+2J_{2}\left(k_{p}R_{p}\right)\right]\frac{k_{p}k_{z}}{kk_{z'}}T_{TM,TE} \right\},
\end{multline}
\begin{multline}
T^{zx}(\mathbf{r},\mathbf{r}') =-\frac{1}{4\pi}\int dk_{p}\,e^{ik_{z}|z|+ik_{z'}|z'|}\\
\times\left\{J_{1}\left(k_{p}R_{p}\right)\frac{k_{p}^{2}}{kk'}T_{TM,TM}\right\},
\end{multline}
for the transmissive part.

%%%%%%%%%%%%%%%%%%%%%%%%%%%%%%%%%%%%%%%%%%%%%%%%%%%%%%%%%%%%%%%%%%%%%%

\end{document}